\pdfoutput=1
\RequirePackage{ifpdf}
\ifpdf 
\documentclass[pdftex]{sigma}
\else
\documentclass{sigma}
\fi

\numberwithin{equation}{section}

\begin{document}

\newcommand{\arXivNumber}{1503.07747}

\allowdisplaybreaks

\renewcommand{\thefootnote}{$\star$}

\renewcommand{\PaperNumber}{061}

\FirstPageHeading

\ShortArticleName{Conf\/luent Chains of DBT: Enlarged Shape Invariance and New Orthogonal Polynomials}

\ArticleName{Conf\/luent Chains of DBT: Enlarged Shape Invariance and New Orthogonal Polynomials\footnote{This paper is a~contribution to the Special Issue on Exact Solvability and Symmetry Avatars
in honour of Luc Vinet.
The full collection is available at
\href{http://www.emis.de/journals/SIGMA/ESSA2014.html}{http://www.emis.de/journals/SIGMA/ESSA2014.html}}}

\Author{Yves GRANDATI~$^\dag$ and Christiane QUESNE~$^\ddag$}

\AuthorNameForHeading{Y.~Grandati and C.~Quesne}

\Address{$^\dag$~Equipe BioPhysStat, LCP A2MC, Universit\'e de Lorraine-Site de Metz,\\
\hphantom{$^\dag$}~1 bvd D.F.~Arago, F-57070, Metz, France}
\EmailD{\href{mailto:grandati@yahoo.fr}{grandati@yahoo.fr}}

\Address{$^\ddag$~Physique Nucl\'eaire Th\'eorique et Physique Math\'ematique,  Universit\'e Libre de
Bruxelles, \\
\hphantom{$^\ddag$}~Campus de la Plaine CP229, Boulevard~du Triomphe, B-1050 Brussels,
Belgium}
\EmailD{\href{mailto:cquesne@ulb.ac.be}{cquesne@ulb.ac.be}}

\ArticleDates{Received March 26, 2015, in f\/inal form July 15, 2015; Published online July 28, 2015}

\Abstract{We construct rational extensions of the Darboux--P\"{o}schl--Teller and
isotonic potentials via two-step conf\/luent Darboux transformations. The former are strictly isospectral to the initial potential, whereas the latter are only quasi-isospectral. Both are
associated to new families of orthogonal polynomials, which, in the f\/irst
case, depend on a continuous parameter. We also prove that these extended
potentials possess an enlarged shape invariance property.}

\Keywords{quantum mechanics; supersymmetry; orthogonal polynomials}

\Classification{81Q05; 81Q60; 42C05}

\renewcommand{\thefootnote}{\arabic{footnote}}
\setcounter{footnote}{0}

\section{Introduction}

Since the seminal paper of G\'omez-Ullate, Kamran, and Milson \cite{gomez3},
which introduced the concept of exceptional orthogonal polynomials (EOP),
the discovery of their connection with translationally shape invariant
quantum potentials (TSIP) by Quesne \cite{quesne1,quesne}, and the construction of inf\/inite sets of such potentials by Odake and Sasaki \cite{odake}, much
progress has been made in the understanding of exactly solvable systems
related to orthogonal polynomials (see \cite{GGM} and re\-fe\-ren\-ces therein).
The key tool to generate such systems is the Darboux or Darboux--B\"{a}cklund
transformation (DBT), which connects pairs of intertwined Hamiltonians.
Starting from one primary TSIP,
specif\/ic symmetries of this last select the quasi-polynomial formal
eigenfunctions that can be used as seed functions to build chains of
rationally extended potentials. The eigenstates of these extensions are then
(up to a gauge factor) exceptional orthogonal polynomials, which, by using
Crum formulas, can be expressed as Wronskians of classical orthogonal
polynomials. The regularity properties of the chains, including degenerate
chains (i.e., chains with repeated use of the same seed functions), are
controlled by enlarged versions of the Krein--Adler theorem \cite{adler,Duran,GGM,krein,samsonov}. For some chains, the extended
potentials share the same shape inva\-rian\-ce properties as the primary
potential \cite{grandati4,grandati5,quesne,quesne4}. With other choices of
seed functions, the resulting potentials possess an enlarged shape
invariance property \cite{grandati6,quesne6,quesne7}.

 Until now, the chains
of extensions were ``rigid'' in the sense that they were uniquely determined by
the tuple of associated seed functions. Very recently, with B.~Bagchi~\cite{BGQ}, we obtained new rational extensions of the Darboux--P\"{o}schl--Teller
potential (based on the so-called para-Jacobi polynomials~\cite{calogero}),
which depend on a free parameter and can then be modulated conti\-nuous\-ly. The
eigenstates of these extended potentials are associated to new families of
orthogonal polynomials that are, in a broad sense, exceptional para-Jacobi
polynomials and which depend on a free continuous parameter.

In this paper, we consider the possibility of building new rational extensions
of two conf\/ining TSIP, namely the trigonometric Darboux--P\"{o}schl--Teller
(TDPT) and isotonic potentials, via conf\/luent chains of DBT, that is chains
of DBT in which the spectral parameters of the dif\/ferent seed functions
converge to the same value. It has to be noticed that it is precisely by using such conf\/luent chains applied to the constant potential and considering the associated rational extensions that Adler and Moser built the Burchnall--Chaundy polynomials~\cite{burchnall} in their seminal paper on the rational solutions of the KdV equation~\cite{adler-moser}.

The possibility of considering two successive factorization energies tending towards a common real value in a chain of Darboux transformations was f\/irst considered in the framework of phase-equivalent potential construction \cite{baye}, then extended to a class of potentials def\/ined on the line~\cite{sparenberg}. These approaches generalized to arbitrary bound-state energies a procedure already known for the ground state and the f\/irst few excited states \cite{cooper, keung}. An independent proposal was made wherein the terminology ``conf\/luent'' algorithm was introduced and some partners of the free particle and the harmonic oscillator were exhibited \cite{mielnik}. The conf\/luent algorithm was then studied in more general terms and applied to the free particle, one-soliton well, and harmonic oscillator \cite{fernandez03}. It was also considered in a general construction of all possible f\/irst- and second-order partners of the TDPT potential \cite{contreras}. The ``hyperconf\/luent'' third-order algorithm, wherein the three factorization energies converge to the same value, was analyzed~\cite{fernandez3}. Some Wronskian formulas applicable to the conf\/luent case were also derived~\cite{fernandez2, fernandez, schulze}.

The present work dif\/fers from the previous ones devoted to conf\/luent chains by the restriction to f\/inal potentials that are rational extensions of the initial ones. As a consequence of this condition, some parameters of the latter may have to be chosen integer.

After recalling in Section~\ref{section2} the basic elements concerning the Darboux--B\"{a}cklund transformations, we review in Section~\ref{section3} the concept
of conf\/luent chains of DBT. We show in particular that the conf\/luent chains
of arbitrary order can be generated within the standard frame of (completed)
DBT chains, giving rise to multiparameter dependent extensions.

In Section~\ref{section4}, applying two-step conf\/luent chains of DBT for which
the seed functions are eigenstates, we build regular rational extensions of
the TDPT with appropriate parameters. The extended potentials depend on a
continuous parameter and are strictly isospectral to the initial potential.
The eigenstates form new families of orthogonal polynomials, which have a~free parameter dependence. We exhibit particular examples and prove that the
extended potentials present an enlarged shape invariance property, in which
the parameter transformation acts in a~nontrivial way on the supplementary
parameter.

In Section~\ref{section5}, we make the same construction for the isotonic system.
In contrast with the TDPT case, the regular rational extensions do not depend on
any supplementary degree of freedom and we only have quasi-isospectrality
between the extended potentials and the original one. We also furnish
explicit examples of extensions and establish their enlarged shape
invariance property.
Section~\ref{section6} contains some f\/inal comments.

\section{Darboux--B\"{a}cklund transformations: basic elements}\label{section2}

We consider a one-dimensional Hamiltonian $\widehat{H}=-d^{2}/dx^{2}+V(x)$, $x\in I\subset \mathbb{R}$, and the associated Schr\"{o}dinger equation
\begin{gather}
  \psi _{\lambda }^{\prime \prime }(x)+ ( E_{\lambda }-V(x) )
  \psi_{\lambda }(x)=0,  \label{EdS1}
\end{gather}
$\psi _{\lambda }(x)$ being a formal eigenfunction of $\widehat{H}$ for the
eigenvalue $E_{\lambda }$. In the following, we suppose that, with Dirichlet
boundary conditions on $I$, $\widehat{H}$ admits a discrete spectrum of
energies and eigenstates $( E_{n},\psi _{n}) _{n\in \{
0,\dots ,n_{\max }\} \mathbb{\subseteq N}}$, where, without loss of
generality, we can always suppose that the ground level of $\widehat{H}$ is
at zero ($E_{0}=0$).

The Riccati--Schr\"{o}dinger (RS) function $w_{\lambda }(x)=-\psi _{\lambda
}^{\prime }(x)/\psi _{\lambda }(x)$ associated to $\psi _{\lambda }$
satisf\/ies the corresponding Riccati--Schr\"{o}dinger  equation~\cite{grandati}
\begin{gather}
  -w_{\lambda }^{\prime }(x)+w_{\lambda }^{2}(x)=V(x)-E_{\lambda }.
\label{edr4}
\end{gather}

From any solution $\psi _{\nu }$ (or equivalently $w_{\nu }$), we can build a
Darboux--B\"{a}cklund transformation (DBT) $A( w_{\nu }) $
def\/ined as~\cite{Ramos,carinena2,darboux2,darboux,grandati}
\begin{gather}
    w_{\lambda }(x)\overset{A ( w_{\nu } ) }{\rightarrow }w_{\lambda
      }^{ ( \nu  ) }(x)=-w_{\nu }(x)+ ( E_{\lambda }-E_{\nu } )
      / ( w_{\nu }(x)-w_{\lambda }(x) ),\nonumber \\
    \psi _{\lambda }(x)\overset{A ( w_{\nu } ) }{\rightarrow }\psi
      _{\lambda }^{ ( \nu  ) }(x)=\exp \left( -\int dxw_{\lambda }^{(\nu
      )}(x)\right) \sim \widehat{A} ( w_{\nu } ) \psi _{\lambda }(x)
 ,\qquad \lambda \neq \nu ,  \label{transfoback2}
\end{gather}
where $\widehat{A} ( w_{\nu } ) $ is a f\/irst-order dif\/ferential
operator given by
\begin{gather*}
  \widehat{A} ( w_{\nu } ) =d/dx+w_{\nu }(x). 
\end{gather*}

$\psi _{\lambda }^{ ( \nu  ) }$ and $w_{\lambda }^{ ( \nu
 ) }$ are respectively solutions of the Schr\"{o}dinger and RS
equations with the same energy $E_{\lambda }$ as in equations~(\ref{EdS1}) and (\ref{edr4}), but with a modif\/ied potential
\begin{gather}
  V^{( \nu) }(x)=V(x)+2w_{\nu }^{\prime }(x),  \label{pottrans}
\end{gather}
that we call an extension of $V(x)$. For the associated Hamiltonian $\widehat{H}^{( \nu) }=-d^{2}/dx^{2}+V^{( \nu) }(x)$,
we have the factorizations
\begin{gather*}
    \widehat{H}^{(\nu) }=\widehat{A} ( w_{\nu } ) \widehat{%
      A}^{+} (w_{\nu }) +E_{\nu }, \qquad
    \widehat{H}=\widehat{A}^{+}( w_{\nu }) \widehat{A} ( w_{\nu
      } ) +E_{\nu },
\end{gather*}
with
\begin{gather*}
  \psi _{\lambda }(x)\sim \widehat{A}^{+}( w_{\nu }) \psi _{\lambda
  }^{( \nu ) }(x).  
\end{gather*}

The function $\psi _{\lambda }^{( \nu) }$ in equation~(\ref{transfoback2}) can then be rewritten as the Darboux--Crum formula
\begin{gather}
  \psi _{\lambda }^{( \nu ) }(x)\sim \frac{W( \psi _{\nu
  },\psi _{\lambda }\,|\, x) }{\psi _{\nu }(x)},  \label{foDBTwronsk}
\end{gather}
where $W( y_{1},\dots ,y_{m}\,|\, x) $ denotes the Wronskian of the
family of functions $y_{1},\dots ,y_{m}$,
\begin{gather*}
  W( y_{1},\dots ,y_{m}\,|\, x) =\left\vert
    \begin{matrix}
    y_{1} ( x ) & \dots  & y_{m}(x) \\
    \dots  &  & \dots  \\
    y_{1}^{( m-1) }(x) & \dots  & y_{m}^{( m-1)}(x)
    \end{matrix}
  \right\vert . 
\end{gather*}

The eigenfunction $\psi _{\nu }$ is called the seed function of the DBT $A(w_{\nu })$ and $V^{( \nu ) }$ and $\psi _{\lambda }^{( \nu
) }$ are the Darboux transforms of $V$ and $\psi _{\lambda }$,
respectively.

Note that $A(w_{\nu })$ annihilates $\psi _{\nu }$ and, consequently, equations~(\ref{transfoback2}) and~(\ref{foDBTwronsk}) allow to obtain an
eigenfunction of $V^{( \nu ) }$ for the eigenvalue $E_{\lambda }$
only when $\lambda \neq \nu $. Nevertheless, we can readily verify that $1/\psi _{\nu }(x)$ is such an eigenfunction. By extension, we then def\/ine
the ``image'' by $A(w_{\nu })$ of the seed eigenfunction~$\psi _{\nu }$ itself
as
\begin{gather}
  \psi _{\nu }^{( \nu) }(x)\sim \frac{1}{\psi _{\nu }(x)}.  \label{psinunu}
\end{gather}

At the formal level, the DBT can be straightforwardly iterated and a chain
of~$m$ DBT can be simply described by the following scheme
\begin{gather*}
    \psi _{\lambda }\overset{A(w_{\nu _{1}})}{\rightarrowtail }\psi _{\lambda
      }^{( \nu _{1}) }\overset{A\big(w_{\nu _{2}}^{( N_{1}) }\big)}{
      \rightarrowtail }\psi _{\lambda }^{( N_{2}) }\cdots \overset{A\big(w_{\nu
      _{m}}^{( N_{m-1}) }\big)}{\rightarrowtail }\psi _{\lambda }^{(
      N_{m}) }, \\
    V\overset{A(w_{\nu _{1}})}{\rightarrowtail }V^{( \nu _{1}) }
      \overset{A\big(w_{\nu _{2}}^{( N_{1}) }\big)}{\rightarrowtail }V^{(
      N_{2}) }\cdots \overset{A\big(w_{\nu _{m}}^{( N_{m-1}) }\big)}{
      \rightarrowtail }V^{( N_{m}) },
\end{gather*}
where $N_{j}$ denotes the $j$-uple $( \nu _{1},\dots ,\nu _{j}) $
(with $N_{1}=\nu _{1}$), which completely characterizes the chain. We denote by $
( N_{m},\nu _{m+1},\dots ,\nu _{m+k}) $ the chain obtained by adding
to the chain $N_{m}$ the DBT associated to the successive eigenfunctions $
\psi _{\nu _{m+1}}^{( N_{m}) },\ldots,\psi _{\nu _{m+k}}^{(N_{m+k-1}) }$.

$\psi _{\lambda }^{( N_{m}) }$ is an eigenfunction associated to
the eigenvalue $E_{\lambda }$ of the potential (see equation~(\ref{pottrans}))
\begin{gather}
  V^{( N_{m}) }(x)=V(x)+2\sum_{j=1}^{m}\big( w_{\nu _{j}}^{(
  N_{j-1}) }(x)\big) ^{\prime }  \label{potnstep}
\end{gather}
and can be written as (cf.\ equations~(\ref{transfoback2}) and (\ref{foDBTwronsk}))
\begin{gather}
  \psi _{\lambda }^{( N_{m}) }(x)=\widehat{A}\big( w_{\nu
  _{m}}^{( N_{m-1}) }\big) \psi _{\lambda }^{(
  N_{m-1}) }(x)=\widehat{A}\big( w_{\nu _{m}}^{( N_{m-1})
  }\big) \cdots \widehat{A}( w_{\nu _{1}}) \psi _{\lambda }(x).
  \label{etats n}
\end{gather}

A chain is non-degenerate if all the spectral indices $\nu _{i}$ of the
chain $N_{m}$ are distinct and is degenerate if some of them are repeated in
the chain. For non-degenerate chains, Crum has derived very useful formulas
for the extended potentials and their eigenfunctions in terms of Wronskians
of eigenfunctions of the initial potential~\cite{crum}.

\textbf{Crum's formulas.}
\textit{When all the  $\nu _{j}$ and  $\lambda $ are
distinct, we have
\begin{gather}
  \psi _{\lambda }^{( N_{m}) }(x)=\frac{W^{( N_{m},\lambda
 ) }(x) }{W^{( N_{m}) }(x) }
  \label{etats n3}
\end{gather}
and
\begin{gather}
  V^{( N_{m}) }(x)=V(x)-2\big( \log W^{( N_{m}) }(
  x) \big) ^{\prime \prime },  \label{potnstep2}
\end{gather}
where  $W^{( N_{m}) }(x) =W( \psi _{\nu
_{1}},\dots ,\psi _{\nu _{m}}\,|\, x) $.}

\section{Conf\/luent chains of DBT}
\label{section3}

The single-conf\/luent limit of a chain of DBT $N_{m}$ is obtained when all
the spectral indices $\nu _{j}$ tend simultaneously to the same value $\nu
_{j}\rightarrow \nu $, $\forall\,  j\in  \{ 1,\dots ,m \} $ (in the
following, we consider only single-conf\/luent chains and than omit the
adjective ``single'').

\subsection{Two-step conf\/luent chains}

We consider a chain of two DBT $N_{2}=( \nu _{1},\nu _{2}) $,
which, in the non-degenerate case $\nu _{1}\neq \nu _{2}$, gives (see equations~(\ref{potnstep}), (\ref{etats n}), (\ref{etats n3}), and~(\ref{potnstep2}))
\begin{gather}
 V^{( \nu _{1},\nu _{2}) }(x) =V(x)
      -2\big( \log W^{ ( \nu _{1},\nu _{2} ) }(x)\big) ^{\prime \prime}    =V(x) -2\big[  ( E_{\nu _{2}}-E_{\nu _{1}} ) / (
      w_{\nu _{2}}(x)-w_{\nu _{1}}(x) ) \big] ^{\prime }, \nonumber\\
 \psi _{\nu _{2}}^{ ( \nu _{1} ) }(x)= ( w_{\nu _{1}}(x)-w_{\nu
      _{2}}(x) ) \psi _{\nu _{2}}(x)=W(\psi _{\nu _{1}},\psi _{\nu _{2}}\,|\,
      x)/\psi _{\nu _{1}}(x).
 \label{1steppsi}
\end{gather}

Note that in the degenerate case $\nu _{1}=\nu _{2}$, we have (see equation~(\ref{psinunu}))
\begin{gather*}
      \psi _{\nu _{1}}^{( \nu _{1}) }(x)=1/\psi _{\nu _{1}}(x), \qquad
    w_{\nu _{1}}^{( \nu _{1}) }(x)=-w_{\nu _{1}}(x),
\end{gather*}
and by applying the DBT $A\big(w_{\nu _{1}}^{( \nu _{1}) }\big)=A(-w_{\nu
_{1}})$ to $V^{( \nu _{1}) }$, we recover simply the initial
potential~$V$,
\begin{gather*}
  V^{( \nu _{1},\nu _{1}) }(x) =V^{( \nu
  _{1}) }(x) -2w_{\nu _{1}}^{\prime }(x)=V(x) .
\end{gather*}

The conf\/luent case corresponds to the limit $\nu _{2}\rightarrow \nu _{1}$.
As proven by Fern\'{a}ndez et al.~\cite{fernandez2,fernandez}, the conf\/luent
extended potential and its eigenstates admit the following integral
representations
\begin{gather}
  \widetilde{V}^{( \nu _{1},\nu _{1}) }(x) =V(
  x) -2\left( \log \left( \int_{x_{0}}^{x}dt\psi _{\nu
  _{1}}^{2}(t)-W_{0}\right) \right) ^{\prime \prime }  \label{confluentpot3}
\end{gather}
and
\begin{gather}
  \widetilde{\psi }_{k}^{( \nu _{1},\nu _{1}) }(x)
  =( E_{\nu _{1}}-E_{k}) \psi _{k}(x) -\frac{\psi _{\nu
  _{1}}^{2}(x)}{\int_{x_{0}}^{x}dt\psi _{\nu _{1}}^{2}(t)-W_{0}}\psi
  _{k}^{( \nu _{1}) }(x) .  \label{confluentstates3}
\end{gather}

Both depend on an arbitrary real parameter~$W_{0}$ and for an adapted
range of $W_{0}$ values, the extended potential is regular. In fact, the
formula for the potential~(\ref{confluentpot3}) already appears in many previous works, for instance in
a 1986 paper of Luban and Pursey~\cite{luban} and a few years later in~\cite{keung}.
The Matveev formulas~\cite{grandati2,matveev,matveev2} for the two-step case
\begin{gather}
    \widetilde{V}^{( \nu _{1},\nu _{1}) }(x) =V(
      x) -2\left[ \log W\left( \psi _{\nu _{1}},\left( \frac{\partial \psi
      _{\nu }(x)}{\partial E_{\nu }}\right) _{\nu =\nu _{1}}\,|\, x\right) \right]
      ^{\prime \prime }, \nonumber\\
    \widetilde{\psi }_{k}^{\left( \nu _{1},\nu _{1}\right) }(x)
      =W\left( \psi _{\nu _{1}},\left( \frac{\partial \psi _{\nu }(x)}{\partial
      E_{\nu }}\right) _{\nu =\nu _{1}},\psi _{k}\,|\, x\right) \Big/W\left( \psi _{\nu
      _{1}},\left( \frac{\partial \psi _{\nu }(x)}{\partial E_{\nu }}\right) _{\nu
      =\nu _{1}}\,|\, x\right) ,
 \label{2stepMatveev}
\end{gather}
which express the conf\/luent extension and its eigenstates in terms of
generalized Wronskians (which are in fact two-way, or double Wronskians~\cite{vein}) can be viewed as associated to a particular choice of the $W_{0}$
constant. Indeed, if we consider the indexed family of RS functions $w_{\nu }(x)
$ as satisfying a prescribed initial condition in $x_{0}$, we have, in the
conf\/luent limit, $w_{\nu _{2}}(x)\rightarrow w_{\nu _{1}}(x)$. It results
from equation~(\ref{1steppsi}) that
\begin{gather*}
  V^{( \nu _{1},\nu _{1}) }(x) =V(x)
  -2\left( 1\Big/\left( \frac{\partial w_{\nu }(x)}{\partial E_{\nu }}\right)
  _{\nu =\nu _{1}}\right) ^{\prime }.  \label{confluentpot}
\end{gather*}

But we can readily verify that
\begin{gather*}
  1/\left( \frac{\partial w_{\nu }(x)}{\partial E_{\nu }}\right) _{\nu =\nu
  _{1}}=\left[ \log W\left( \psi _{\nu _{1}},\left( \frac{\partial \psi _{\nu
  }(x)}{\partial E_{\nu }}\right) _{\nu =\nu _{1}}\,|\, x\right) \right]
  ^{\prime }
\end{gather*}
and since in this case we also have \cite{messiah}
\begin{gather*}
  \left( \frac{\partial w_{\nu }(x)}{\partial E_{\nu }}\right) _{\nu =\nu
  _{1}}=\frac{1}{\psi _{\nu _{1}}^{2}(x)}\int_{x_{0}}^{x}dt\psi _{\nu
  _{1}}^{2}(t),  
\end{gather*}
we see that equation~(\ref{2stepMatveev}) corresponds to equations~(\ref{confluentpot3}) and
(\ref{confluentstates3}) with $W_{0}=0$.

It has to be noticed that the degenerate extension $V^{( \nu _{1},\nu
_{1}) }(x) $ can be recovered from\ the conf\/luent one $\widetilde{V}^{( \nu _{1},\nu _{1}) }(x) $ by taking
the singular limit value $W_{0}\rightarrow \infty $.

Fern\'{a}ndez et al.\ used these formulas to generate new second-order SUSY
partners of the free particle, the Kepler--Coulomb, and the single-gap Lam\'{e}
potentials~\cite{fernandez2, fernandez}.

The preceding results can be in fact integrated within the standard DBT
scheme simply using the DBT in its completed form (see equation~(\ref{psinunu}))
as in~\cite{cooper,keung}. Indeed, by applying the DBT~$A(w_{\nu })$,
we generate f\/irst the one-step (possibly singular) extension
\begin{gather*}
  V^{( \nu ) }(x)=V(x)+2w_{\nu }^{\prime }(x)=V(x)-2( \log
  \psi _{\nu }(x)) ^{\prime \prime }.
\end{gather*}

Since $\psi _{\nu }^{(\nu) }=1/\psi _{\nu }$ is an
eigenfunction of $V^{(\nu) }$ for the eigenvalue $E_{\nu }$,
the most general eigenfunction (up to a multiplicative factor) of $V^{(\nu) }$ for the same eigenvalue is
\begin{gather}
  \Psi _{\nu }^{(\nu) }(x;\lambda _{1})=\psi _{\nu }^{( \nu
) }(x)\left( \lambda _{1}+\int_{x_{0}}^{x}dt\frac{1}{( \psi _{\nu
  }^{(\nu) }(t)) ^{2}}\right) =\frac{\lambda
  _{1}+\int_{x_{0}}^{x}dt\psi _{\nu }^{2}(t)}{\psi _{\nu }(x)},\qquad \lambda
  _{1}\in \mathbb{R},  \label{2stepgenseed}
\end{gather}
and the corresponding RS function is
\begin{gather*}
  W_{\nu }^{(\nu) }(x;\lambda _{1})=-\left[ \log \left( \frac{
  \lambda _{1}+\int_{x_{0}}^{x}dt\psi _{\nu }^{2}(t)}{\psi _{\nu }(x)}\right)
  \right] ^{\prime }=-w_{\nu }(x)-\frac{\psi _{\nu }^{2}(x)}{\lambda
  _{1}+\int_{x_{0}}^{x}dt\psi _{\nu }^{2}(t)}.
\end{gather*}

We can now use this general solution as seed function for the second DBT.
Then ap\-plying~$A\big(W_{\nu }^{(\nu) }\big)$ to~$V^{(\nu)} $, we obtain the following second extension
\begin{gather}
  \widetilde{V}^{ ( \nu ,\nu  ) }(x;\lambda _{1})  =V^{ ( \nu
     ) }(x)-2\big( \log \Psi _{\nu }^{(\nu) }(x;\lambda
    _{1})\big) ^{\prime \prime }=V(x)-2\big[ \log \big( \psi _{\nu }(x)\Psi
    _{\nu }^{(\nu) }(x;\lambda _{1})\big) \big] ^{\prime \prime
    }  \nonumber\\
    \hphantom{\widetilde{V}^{ ( \nu ,\nu  ) }(x;\lambda _{1})}{}
 =V(x)-2\left[ \log \left( \lambda _{1}+\int_{x_{0}}^{x}dt\psi _{\nu
    }^{2}(t)\right) \right] ^{\prime \prime }
 =V(x)-2\left( \frac{\psi _{\nu
    }^{2}(x)}{\lambda _{1}+\int_{x_{0}}^{x}dt\psi _{\nu }^{2}(t)}\right)
    ^{\prime }\!, \!\!\! \label{2stepgenform}
\end{gather}
and we recover the f\/irst Fern\'{a}ndez formula (\ref{confluentpot3}) with $
\lambda _{1}=-W_{0}$. As for the eigenfunctions of~$\widetilde{V}^{(
\nu ,\nu ) }$, they are given by~($\mu \neq \nu $)
\begin{gather}
  \widetilde{\psi }_{\mu }^{( \nu ,\nu ) }(x;\lambda _{1})=\widehat{
  A}\big(W_{\nu }^{(\nu) }\big)\psi _{\mu }^{(\nu)
  }(x)=( E_{\nu }-E_{\mu }) \psi _{\mu}(x) -\frac{
  W(\psi _{\nu },\psi _{\mu }\,|\, x)}{\Psi _{\nu }^{(\nu)
  }(x;\lambda _{1})}.  \label{2stepgeneig2}
\end{gather}

\subsection{General multi-step conf\/luent chains}

Fern\'{a}ndez and Salinas-Hern\'{a}ndez \cite{fernandez3} have also
considered the so-called ``hyperconf\/luent'' case corresponding to a three-step
conf\/luent DBT, for which they have extended the previous formulas, equations~(\ref{confluentpot3}) and (\ref{confluentstates3}). For these three-step
extensions, the potential depends on two arbitrary real parameters.

In fact, the preceding analysis allows to obtain integral formulas ``\`{a} la
Fern\'{a}ndez'' for chains of arbitrary order in a very simple way. In the
following, the symbol $( \nu ^{l}) $ means $\underset{l~\text{times}}{\underbrace{( \nu ,\dots ,\nu ) }}$.

In the three-step case, the image of $\Psi _{\nu }^{(\nu) }$
(see equation~(\ref{2stepgenseed})) by the DBT $A(W_{\nu }^{(\nu) })$
being $\psi _{\nu }^{( \nu ^{2}) }=1/\Psi _{\nu }^{( \nu
) }$, the general eigenfunction of~$V^{( \nu ^{2})
}(x;\lambda _{1})$ (see equation~(\ref{2stepgenform})) associated to the eigenvalue
$E_{\nu }$ is
\begin{gather*}
  \Psi _{\nu }^{( \nu ^{2}) }(x;\Lambda _{2})=\frac{\lambda
  _{2}+\int_{x_{0}}^{x}dt\big( \Psi _{\nu }^{(\nu) }(t;\lambda
  _{1})\big) ^{2}}{\Psi _{\nu }^{(\nu) }(x;\lambda _{1})},\qquad
  \lambda _{1},\lambda _{2}\in \mathbb{R},
\end{gather*}
where we have used the notation $\Lambda _{m}=  (\lambda
_{1},\dots ,\lambda _{m} ) $. The next extension generated by the DBT $A\bigl(W_{\nu }^{( \nu ^{2}) }\bigr)$ is then
\begin{gather*}
  \widetilde{V}^{( \nu ^{3}) }(x;\Lambda _{2}) =V(x)-2\big[ \log
    \big( \psi _{\nu }(x)\Psi _{\nu }^{(\nu) }(x;\lambda _{1})\Psi
    _{\nu }^{( \nu ^{2}) }(x;\Lambda _{2})\big) \big] ^{\prime
    \prime } \\
\hphantom{\widetilde{V}^{( \nu ^{3}) }(x;\Lambda _{2})}{}
=V(x)-2 ( \log \psi _{\nu }(x) ) ^{\prime \prime }-2\left(
    \lambda _{2}+\int_{x_{0}}^{x}dt\big( \Psi _{\nu }^{(\nu)
    }(t;\lambda _{1})\big) ^{2}\right) ^{\prime \prime }.
\end{gather*}

We recover the ``hyperconf\/luent'' third-order superpartner of $V(x)$ as
obtained by Fern\'{a}ndez and Salinas-Hern\'{a}ndez~\cite{fernandez3}.
Within this scheme, the generalization is immediate and re\-pea\-ting the
procedure~$m$ times, we obtain for the hyperconf\/luent $m^{\rm th}$-order
extension of~$V(x)$ the expression
\begin{gather*}
  \widetilde{V}^{( \nu ^{m}) }(x;\Lambda _{m-1}) =V(x)-2\left( \log
    \prod\limits_{j=0}^{m-1}\Psi _{\nu }^{( \nu ^{j}) }(x;\Lambda
    _{j})\right) ^{\prime \prime }
    =V(x)-2\sum_{j=0}^{m-1}\big( \log \Psi _{\nu
  }^{( \nu ^{j}) }(x;\Lambda _{j})\big) ^{\prime \prime },
\end{gather*}
with the following recurrence relation for the successive seed functions
\begin{gather*}
  \Psi _{\nu }^{( \nu ^{k}) }(x;\Lambda _{k})=\frac{\lambda
  _{k}+{\displaystyle \int}_{x_{0}}^{x}dt\big( \Psi _{\nu }^{( \nu ^{k-1})
  }(t;\Lambda _{k-1})\big) ^{2}}{\Psi _{\nu }^{( \nu ^{k-1})
  }(x;\Lambda _{k-1})}, 
\end{gather*}
where $\Psi _{\nu }^{(0) }(x;\Lambda _{0})=\psi _{\nu }(x)$.

In other words, whenever $m$ is even ($m=2k$)
\begin{gather*}
  \widetilde{V}^{( \nu ^{2k}) }(x;\Lambda _{2k-1})=V(x)-2\left\{ \log
  \left[\prod\limits_{l=0}^{k-1}\left( \lambda _{2l+1}+\int_{x_{0}}^{x}dt\big(
  \Psi _{\nu }^{( \nu ^{2l}) }(t;\Lambda _{2l})\big) ^{2}\right)
  \right]\right\} ^{\prime \prime }  \label{2kstep}
\end{gather*}
and whenever $m$ is odd ($m=2k+1$)
\begin{gather*}
  \widetilde{V}^{( \nu ^{2k+1}) }(x;\Lambda _{2k})=V(x)-2\left\{ \log
  \left[\psi _{\nu }(x)\prod\limits_{l=1}^{k}\left( \lambda
  _{2l}+\int_{x_{0}}^{x}dt\big( \Psi _{\nu }^{( \nu ^{2l-1})
  }(t;\Lambda _{2l-1})\big) ^{2}\right) \right]\right\} ^{\prime \prime }.
\end{gather*}

The eigenstates of $\widetilde{V}^{( \nu ^{m}) }$ can be obtained
by successive applications of the $\widehat{A}\bigl(W_{\nu }^{( \nu
^{l}) }\bigr)$ operators (the product being ordered in decreasing order),
\begin{gather*}
  \widetilde{\psi }_{k}^{( \nu ^{m}) }(x;\Lambda
  _{m-1})=\prod\limits_{l=1}^{m}\widehat{A}\bigl(W_{\nu }^{( \nu
  ^{l-1}) }\bigr)\psi _{k}(x).
\end{gather*}

A direct application of the Crum Wronskian formulas~\cite{crum} to these
general, parameter-dependent, conf\/luent extensions is obviously not possible
and, as mentioned above, the Matveev formulas~\cite{grandati2,matveev,matveev2}
correspond only to a particular choice of the~$\lambda _{j}$ parameters.
Nevertheless, these extended potentials are amenable to other (standard)
Wronskian formulas~\cite{fernandez2,fernandez,schulze}.

In the following, we limit our analysis to the case of two-step DBT. We are
interested in the possibility of building regular and rational extensions with
such conf\/luent chains, which turns out to be possible for the trigonometric
Darboux--P\"{o}schl--Teller (TDPT) potential and the isotonic potential.

\section{Two-step conf\/luent rational extensions of the trigonometric\\
Darboux--P\"{o}schl--Teller (TDPT) potential}
\label{section4}

\subsection{General scheme}

The trigonometric Darboux--P\"{o}schl--Teller (TDPT) potential (with zero
ground-state energy) is def\/ined on$\ x\in {} ] 0,\pi /2[ $ by{\samepage
\begin{gather*}
  V(x;\alpha ,\beta )=\frac{(\alpha +1/2)(\alpha -1/2)}{\sin ^{2}x}+\frac{
  (\beta +1/2)(\beta -1/2)}{\cos ^{2}x}-(\alpha +\beta +1)^{2},  \label{TDPT}
\end{gather*}
with $\alpha ,\beta >1/2$.}

Its physical spectrum, associated to the asymptotic Dirichlet boundary
conditions
\begin{gather*}
  \psi ( 0^{+};\alpha ,\beta ) =0=\psi \left( \left( \frac{\pi }{2}
  \right) ^{-};\alpha ,\beta \right),   
\end{gather*}
is given in terms of Jacobi polynomials \cite{magnus,grandatiDPT,szego}
\begin{gather*}
  \mathit{P}_{n}^{( \alpha ,\beta ) }(z)=\frac{( -1)
  ^{n}\Gamma ( n+\beta +1) }{n!\Gamma ( n+\alpha +\beta
  +1) }\sum_{k=0}^{n}(-1) ^{k}\binom{n}{k}\frac{\Gamma
  ( n+\alpha +\beta +1+k) }{2^{k}\Gamma ( \beta +1+k) }
( 1+z) ^{k},
\end{gather*}
by
\begin{gather*}
E_{n}( \alpha ,\beta ) =(\alpha _{n}+\beta
      _{n}+1)^{2}-(\alpha +\beta +1)^{2}=4n(\alpha +\beta +1+n), \\
    \psi _{n}( x;\alpha ,\beta ) =\psi _{0}( x;\alpha ,\beta
   ) P_{n}^{( \alpha ,\beta ) }(z),
\qquad  n\in \mathbb{N},  
\end{gather*}

with $z=\cos 2x\in {}]{-}1,1[ $, $(\alpha _{n},\beta _{n})=(\alpha
+n,\beta +n)$, and
\begin{gather*}
  \psi _{0}( x;\alpha ,\beta ) =( 1-z) ^{( \alpha
  +1/2) /2}(1+z) ^{( \beta +1/2) /2}.
\end{gather*}

In the following, in order to get rational extensions, we consider the case where~$\alpha $ and~$\beta $ are
integers: $\alpha =N\geq 1$, $\beta =M\geq 1$.

If we choose as initial seed function an eigenstate $\psi _{n}(
x;N,M) $ of $V(x;N,M) $, by taking $x_{0}=\pi /2$ ($z_{0}=-1$), the quantity
\begin{gather*}
  Q_{n}^{(N,M) }(z) =\int_{\pi /2}^{x}d\xi \psi
  _{n}^{2}(\xi ;N,M)=-\frac{1}{2}\int_{-1}^{z}d\zeta (1-\zeta)
  ^{N}(1+\zeta) ^{M}\big( P_{n}^{(N,M) }( \zeta
) \big) ^{2}  
\end{gather*}
is a polynomial of degree $N+M+2n+1$ in $z$ with \cite{magnus,szego}
\begin{gather}
  Q_{n}^{(N,M) }(1)  =-\frac{1}{2}\int_{-1}^{1}d\zeta
    (1-\zeta) ^{N}(1+\zeta) ^{M}\big( P_{n}^{(
    N,M) }(\zeta) \big) ^{2}  \nonumber \\
\hphantom{Q_{n}^{(N,M) }(1)}{}=-\frac{2^{N+M}}{2n+N+M+1}\frac{(n+N) !(n+M) !}{
    n!( n+N+M) !}.  \label{Q1}
\end{gather}

Note the following recurrence
\begin{gather}
  Q_{n-1}^{( N+1,M+1) }(1) =\frac{4n}{n+N+M+1}
  Q_{n}^{(N,M) }(1) .  \label{recQ}
\end{gather}

From equation~(\ref{2stepgenform}), we then obtain for the conf\/luent two-step extension $\widetilde{V}^{(
n^{2}) }$,
\begin{gather*}
  \widetilde{V}^{( n^{2}) }(x;N,M,\lambda _{1}) =V(
    x;N,M) -2\big[ \log \big( \lambda _{1}+Q_{n}^{(N,M)
    }(z) \big) \big] ^{\prime \prime } \\
  \qquad {}= V(x;N,M) +4\big( 1-z^{2}\big) ^{1/2}\frac{d}{dz}\left(
    \frac{(1-z) ^{N+1/2}(1+z) ^{M+1/2}\big(
    P_{n}^{(N,M) }(z) \big) ^{2}}{\lambda
    _{1}+Q_{n}^{(N,M) }(z) }\right) ,
\end{gather*}
where $\widetilde{V}^{( n^{2}) }(x;N,M,\lambda _{1})$ constitutes
a rational extension of $V(x)$ (in the $z$ variable). $Q_{n}^{(
N,M) }(z) $ is strictly decreasing on the interval $]{-}1,1[$ and keeps a negative value, varying from $0$ to $Q_{n}^{(
N,M) }(1)<0$ when~$z$ runs through $]{-}1,1[$.
Consequently, when
\begin{gather}
  \lambda _{1}\in {}] -\infty ,0] \cup {}\big] {-}Q_{n}^{(
  N,M) }(1),+\infty \big[ ,  \label{reglambda}
\end{gather}
then $\lambda _{1}+Q_{n}^{(N,M) }(z) $ keeps a
constant sign, strictly negative or strictly positive respectively, and $\widetilde{V}^{( n^{2}) }(x;N,M,\lambda _{1})$ is regular.

In this case, its eigenfunctions for $k\neq n$ are given by (see equations~(\ref{2stepgenseed})
and~(\ref{2stepgeneig2}))
\begin{gather*}
  \widetilde{\psi }_{k}^{( n^{2}) }(x;N,M,\lambda _{1})=[
  E_{n}(N,M) -E_{k}(N,M) ] \psi _{k}(
  x) -\frac{W(\psi _{n},\psi _{k}\,|\, x)}{\Psi _{n}^{(n)
  }(x;N,M,\lambda _{1})},
\end{gather*}
with
  \begin{gather*}
  \Psi _{n}^{(n) }(x;N,M,\lambda _{1})=(1-z) ^{-(
  N+1/2) /2}(1+z) ^{-( M+1/2) /2}\frac{\lambda
  _{1}+Q_{n}^{(N,M) }(z) }{P_{n}^{(N,M)
  }(z) }  
\end{gather*}
and
\begin{gather*}
  W( \psi _{n},\psi _{k}\,|\, x) =-(1-z) ^{N+1}(
  1+z) ^{M+1}P_{n,k}^{(N,M) }(z) ,
\end{gather*}
where ($P_{-1}^{( \alpha ,\beta ) }(z) =0$)
\begin{gather*}
  P_{n,k}^{(N,M) }(z)  =  ( k+N+M+1 )
    P_{n}^{(N,M) }(z) P_{k-1}^{( N+1,M+1)
    }(z) \\
\hphantom{P_{n,k}^{(N,M) }(z)  =}{} - ( n+N+M+1 ) P_{n-1}^{( N+1,M+1) }(z)
    P_{k}^{(N,M) }(z)
\end{gather*}
is an exceptional Jacobi polynomial in the broad sense of the term.

Hence, for $k\neq n$,
\begin{gather*}
  \widetilde{\psi }_{k}^{( n^{2}) }(x;N,M,\lambda _{1}) =[
    E_{n}(N,M) -E_{k}(N,M) ] \psi _{k}(
    x;N,M) \\
\hphantom{\widetilde{\psi }_{k}^{( n^{2}) }(x;N,M,\lambda _{1}) =}{}+(1-z) ^{( 3N+5/2) /2}(1+z) ^{(
    3M+5/2) /2}\frac{P_{n,k}^{(N,M) }(z)
    P_{n}^{(N,M) }(z) }{\lambda _{1}+Q_{n}^{(
    N,M) }(z) },
\end{gather*}
that is
\begin{gather*}
  \widetilde{\psi }_{k}^{( n^{2}) }(x;N,M,\lambda _{1})=(
  1-z) ^{( N+1/2) /2}(1+z) ^{( M+1/2)
  /2}\frac{\widetilde{P}_{N,M,k}^{( n^{2}) }( z; \lambda_1) }{
  \lambda _{1}+Q_{n}^{(N,M) }(z) },
\end{gather*}
where
\begin{gather*}
  \widetilde{P}_{N,M,k}^{( n^{2}) }( z; \lambda_1) =4(
    n-k) ( n+k+N+M+1) P_{k}^{(N,M) }(z)
    \big[ \lambda _{1}+Q_{n}^{(N,M) }(z) \big] \\
\hphantom{\widetilde{P}_{N,M,k}^{( n^{2}) }( z; \lambda_1) =}{}+(1-z) ^{N+1}(1+z) ^{M+1}P_{n,k}^{(
    N,M) }(z) P_{n}^{(N,M) }(z) .
\end{gather*}

Moreover
\begin{gather*}
  \widetilde{\psi }_{n}^{( n^{2}) }(x;N,M,\lambda _{1}) =1/\Psi
    _{n}^{(n) }(x;N,M,\lambda _{1}) \\
\hphantom{\widetilde{\psi }_{n}^{( n^{2}) }(x;N,M,\lambda _{1})}{}
=(1-z) ^{( N+1/2) /2}(1+z) ^{(M+1/2) /2}\frac{P_{n}^{(N,M) }(z) }{\lambda
    _{1}+Q_{n}^{(N,M) }(z) },
\end{gather*}
which is a normalizable eigenstate of $\widetilde{V}^{(n^{2})
}(x;N,M,\lambda _{1})$. This corresponds to def\/ining
\begin{gather*}
  \widetilde{P}_{N,M,n}^{( n^{2}) }( z; \lambda_1) =P_{n}^{( N,M) }(z) .
\end{gather*}

$\widetilde{\psi }_{k}^{( n^{2}) }(x;N,M,\lambda _{1})$ tends to zero at $z=-1$ and $z=1$ (i.e.,
$x=\pi /2$ and $x=0$) and is then an admissible
eigenstate of $\widetilde{V}^{( n^{2}) }(x;N,M,\lambda _{1})$ for
every $k\geq 0$. The potentials $V(x;N,M) $ and $\widetilde{V}
^{( n^{2}) }(x;N,M,\lambda _{1})$ are therefore strictly
isospectral.

The orthogonality conditions between eigenstates imply that the $\widetilde{P
}_{N,M,k}^{(n^{2})}( z; \lambda_1) $ constitute a~family of
orthogonal polynomials (indexed by $k\in \mathbb{N}$) on $]{-}1,1[ $ with respect to the measure
\begin{gather*}
  \mu _{N,M}^{(n^{2})}(z;\lambda _{1})=\frac{1}{2}\frac{(
  1-z) ^{N}(1+z) ^{M}}{\big( \lambda _{1}+Q_{n}^{( N,M) }(z) \big) ^{2}}.
\end{gather*}

It is worth observing here that the conf\/luent two-step extension $\widetilde{V}^{(
n^{2}) }(x;N,M,\lambda_1)$ may be considered as a special case of one of those with general parameters that have been built
by Contreras-Astorga and Fern\'andez in~\cite[Section~3.2.3(b)]{contreras} (namely the third one given in equation~(3.43)), whenever their parameters~$\lambda$, $\nu$ assume the half-integer values $\lambda = N + 1/2$ and $\nu = M + 1/2$.

\subsection[Examples $n=0$]{Examples $\boldsymbol{n=0}$}

We consider the $n=0$ case. Then
\begin{gather*}
  Q_{0}^{(N,M) }(z)  =-\frac{1}{2}\int_{-1}^{z}d\zeta
    (1-\zeta) ^{N}(1+\zeta) ^{M}
   =- ( z+1 )
    ^{M+1}\sum_{k=0}^{N}\frac{2^{N-k-1}(-1) ^{k}}{M+k+1}\binom{N}{k}
     ( z+1 ) ^{k}
\end{gather*}
and
\begin{gather*}
  \widetilde{V}^{( 0^{2}) }(x;N,M,\lambda _{1}) =V(
    x;N,M) +2\frac{(1-z) ^{2N+1}(1+z) ^{2M+1}}{
    \big( \lambda _{1}+Q_{0}^{(N,M) }(z) \big) ^{2}} \\
\hphantom{\widetilde{V}^{( 0^{2}) }(x;N,M,\lambda _{1}) =}{}
+4(1-z) ^{N}(1+z) ^{M}\frac{M-N-(N+M+1)z}{\lambda
_{1}+Q_{0}^{(N,M) }(z) }.
\end{gather*}

Moreover
\begin{gather*}
  \widetilde{\psi }_{k}^{(0^{2})}(x;N,M,\lambda _{1})=(
  1-z) ^{(N+1/2) /2}(1+z) ^{(M+1/2)
  /2}\frac{\widetilde{P}_{N,M,k}^{(0^{2})}(z; \lambda_1) }{
  \lambda _{1}+Q_{0}^{(N,M) }(z) },
\end{gather*}
where
\begin{gather*}
  \widetilde{P}_{N,M,k}^{(0^{2})}(z; \lambda_1)  = -4k (
     k+N+M+1 ) P_{k}^{(N,M) }(z) \big( \lambda
     _{1}+Q_{0}^{(N,M) }(z) \big) \\
\hphantom{\widetilde{P}_{N,M,k}^{(0^{2})}(z; \lambda_1)  =}{}
+(1-z) ^{N+1}(1+z) ^{M+1}P_{0,k}^{( N,M) }(z) ,  \qquad k\ne 0,
\\
  \widetilde{P}_{N,M,0}^{(0^{2})}(z; \lambda_1) =1,
\end{gather*}
with ($P_{-1}^{(N,M) }(z) =0$)
\begin{gather*}
  P_{0,k}^{(N,M) }(z) =(k+N+M+1)
  P_{k-1}^{(N+1,M+1) }(z) .
\end{gather*}

\subsubsection[The $N=M=1$ case]{The $\boldsymbol{N=M=1}$ case}

The results read
\begin{gather*}
  Q_{0}^{(1,1) }(z) =-\frac{1}{2}(z+1)
  ^{2}\left( 1-\frac{1}{3}(z+1) \right) ,
\\
  \widetilde{V}^{(0^{2})}(x;1,1,\lambda _{1})  = V (
    x;1,1 ) -12\frac{z \big( 1-z^{2} \big) }{\lambda _{1}-\frac{1}{2} (
    z+1 ) ^{2}+\frac{1}{6}(z+1) ^{3}} \\
\hphantom{\widetilde{V}^{(0^{2})}(x;1,1,\lambda _{1})  =}{}+2\frac{\big( 1-z^{2}\big) ^{3}}{\big[ \lambda _{1}-\frac{1}{2}
    (z+1)^{2}+\frac{1}{6}(z+1) ^{3}\big] ^{2}}
\end{gather*}
or
\begin{gather*}
  \widetilde{V}^{(0^{2})}(x;1,1,\lambda _{1})  = \frac{3}{4\sin^2 x}
    + \frac{3}{4\cos^2 x} - 9 - 12 \frac{\sin^2 2x \cos 2x}{\lambda_1 - 2 \cos^4 x
    + \frac{4}{3} \cos^6 x} \\
\hphantom{\widetilde{V}^{(0^{2})}(x;1,1,\lambda _{1})  =}{}+ 2 \frac{\sin^6 2x}{\big(\lambda_1 - 2 \cos^4 x + \frac{4}{3} \cos^6 x
    \big)^2}
\end{gather*}
with $\lambda _{1}\in {}]{-}\infty ,0] \cup {}] \frac{2}{3}, +\infty [$, and
\begin{gather*}
  \widetilde{\psi }_{k}^{(0^{2})}(x;1,1,\lambda _{1})=\left(
  1-z\right) ^{3/4}(1+z) ^{3/4}\frac{\widetilde{P}_{1,1,k}^{\left(
  0^{2}\right) }(z; \lambda_1) }{\lambda _{1}-\frac{1}{2}\left(
  z+1\right) ^{2}+\frac{1}{6}(z+1) ^{3}},
\end{gather*}
where
\begin{gather*}
  \widetilde{P}_{1,1,k}^{(0^{2})}(z; \lambda_1)  = -4k (
    k+3 ) P_{k}^{(1,1) }(z) \left[ \lambda
    _{1}-\frac{1}{2}(z+1) ^{2}+\frac{1}{6}(z+1) ^{3}\right] \\
\hphantom{\widetilde{P}_{1,1,k}^{(0^{2})}(z; \lambda_1)  =}{}+ ( k+3 ) (1-z) ^{2}(1+z)
    ^{2}P_{k-1}^{( 2,2) }(z), \qquad k\ne 0,
\\
  \widetilde{P}_{1,1,0}^{(0^{2})}(z; \lambda_1) = 1.
\end{gather*}

\subsubsection[The $N=2$, $M=1$ case]{The $\boldsymbol{N=2}$, $\boldsymbol{M=1}$ case}

The results read
\begin{gather*}
  Q_{0}^{( 2,1) }(z) =-(z+1) ^{2}\left( 1-
  \frac{2}{3}(z+1) +\frac{1}{8}(z+1) ^{2}\right) ,
\\
  \widetilde{V}^{(0^{2})}(x;2,1,\lambda _{1})  = V (
    x;2,1 ) -4\frac{ ( 1+4z ) (1-z) ^{2} (
    1+z ) }{\lambda _{1}-(z+1) ^{2}+\frac{2}{3} (
    z+1 ) ^{3}-\frac{1}{8}(z+1) ^{4}} \\
\hphantom{\widetilde{V}^{(0^{2})}(x;2,1,\lambda _{1})  =}{}+2\frac{(1-z) ^{5}(1+z) ^{3}}{\big[ \lambda
    _{1}-(z+1) ^{2}+\frac{2}{3}(z+1) ^{3}-\frac{1}{8}
    (z+1) ^{4}\big] ^{2}}
\end{gather*}
or
\begin{gather*}
  \widetilde{V}^{(0^{2})}(x;2,1,\lambda _{1})  = \frac{15}{4\sin^2 x}
    + \frac{3}{4\cos^2 x} - 16 - 32 \frac{(1-4\cos 2x) \sin^4 x \cos^2 x}
    {\lambda_1 - 4\cos^4 x + \frac{16}{3}\cos^6 x - 2\cos^8 x} \\
\hphantom{\widetilde{V}^{(0^{2})}(x;2,1,\lambda _{1})  =} {}+ 512 \frac{\sin^{10} x \cos^6 x}{\big(\lambda_1 - 4\cos^4 x + \frac{16}{3}\cos^6 x
    - 2\cos^8 x\big)^2}
\end{gather*}
with $\lambda _{1}\in {}] {-}\infty ,0 ] \cup {}] \frac{2}{3}, +\infty  [$, and
\begin{gather*}
  \widetilde{\psi }_{k}^{(0^{2})}(x;2,1,\lambda _{1})=(
  1-z) ^{5/4}(1+z) ^{3/4}\frac{\widetilde{P}_{2,1,k}^{(
  0^{2}) }(z; \lambda_1) }{\lambda _{1}-(z+1) ^{2}+\frac{2}{
  3}(z+1) ^{3}-\frac{1}{8}(z+1) ^{4}},
\end{gather*}
where
\begin{gather*}
  \widetilde{P}_{2,1,k}^{(0^{2})}(z; \lambda_1)  = -4k (
    k+4 ) P_{k}^{( 2,1) }(z) \left( \lambda
    _{1}-(z+1) ^{2}+\frac{2}{3}(z+1) ^{3}-\frac{1}{8}
    (z+1) ^{4}\right) \\
\hphantom{\widetilde{P}_{2,1,k}^{(0^{2})}(z; \lambda_1)  =} {}+( k+4) (1-z) ^{3}(1+z)
    ^{2}P_{k-1}^{( 3,2) }(z), \qquad k\ne 0,
\\
  \widetilde{P}_{2,1,0}^{(0^{2})}(z; \lambda_1) = 1.
\end{gather*}

\subsubsection[The $N=1$, $M=2$ case]{The $\boldsymbol{N=1}$, $\boldsymbol{M=2}$ case}

The results read
\begin{gather*}
  Q_{0}^{(1,2) }(z) =-(z+1) ^{3}\left(
  \frac{1}{3}-\frac{1}{8}(z+1) \right) ,
\\
  \widetilde{V}^{(0^{2})}(x;1,2,\lambda _{1})  = V (
    x;1,2 ) +4\frac{( 1-4z) (1-z) (1+z)
    ^{2}}{\lambda _{1}-\frac{1}{3}(z+1) ^{3}+\frac{1}{8}(
    z+1) ^{4}} \\
\hphantom{\widetilde{V}^{(0^{2})}(x;1,2,\lambda _{1})  =}{}+2\frac{(1-z) ^{3}(1+z) ^{5}}{\big[ \lambda _{1}-
    \frac{1}{3}(z+1) ^{3}+\frac{1}{8}(z+1) ^{4}\big]^{2}}
\end{gather*}
or
\begin{gather*}
  \widetilde{V}^{(0^{2})}(x;1,2,\lambda _{1})  = \frac{3}{4\sin^2 x}
    + \frac{15}{4\cos^2 x} - 16 + 32 \frac{(1-4\cos 2x) \sin^2 x \cos^4 x}
    {\lambda_1 - \frac{8}{3}\cos^6 x + 2\cos^8 x} \\
\hphantom{\widetilde{V}^{(0^{2})}(x;1,2,\lambda _{1})  =} {}+ 512 \frac{\sin^6 x \cos^{10}x}{\big(\lambda_1 - \frac{8}{3}\cos^6 x +
    2\cos^8 x\big)^2}
\end{gather*}
with $\lambda _{1}\in {}]{-}\infty ,0 ] \cup {}] \frac{2}{3}, +\infty[$, and
\begin{gather*}
  \widetilde{\psi }_{k}^{(0^{2})}(x;1,2,\lambda _{1})=(
  1-z) ^{3/4}(1+z) ^{5/4}\frac{\widetilde{P}_{1,2,k}^{(
  0^{2}) }(z; \lambda_1) }{\lambda _{1}-\frac{1}{3}(z+1)
  ^{3}+\frac{1}{8}(z+1) ^{4}},
\end{gather*}
where
\begin{gather*}
  \widetilde{P}_{1,2,k}^{(0^{2})}(z; \lambda_1)  = -4k (
    k+4 ) P_{k}^{(1,2) }(z) \left( \lambda _{1}-
    \frac{1}{3}(z+1) ^{3}+\frac{1}{8}(z+1) ^{4}\right) \\
\hphantom{\widetilde{P}_{1,2,k}^{(0^{2})}(z; \lambda_1)  =}{}+(k+4) (1-z) ^{2}(1+z)
    ^{3}P_{k-1}^{(2,3) }(z), \qquad k\ne 0,
\\
  \widetilde{P}_{1,2,0}^{(0^{2})}(z; \lambda_1) = 1.
\end{gather*}

\subsection{Shape invariance of the two-step conf\/luent rational extensions\\
of the TDPT potential}

$V(x;N,M) $ is a translationally shape invariant potential with
a SUSY partner
\begin{gather*}
  V_{\rm SUSY}(x;N,M)=V^{(0) }(x;N,M) =V(
  x;N+1,M+1) +E_{1}(N,M)
\end{gather*}
and (the coef\/f\/icient is readily established by a direct calculation)
\begin{gather*}
  \psi _{n}^{(0) }(x;N,M) =\frac{W(\psi _{0},\psi
  _{n}\,|\, x)}{\psi _{0}(x;N,M) }=\left( -\frac{E_{n}(
  N,M) }{4n}\right) \psi _{n-1}( x;N+1,M+1) .
\end{gather*}

Since $\widetilde{V}^{(n^{2})}(x;N,M,\lambda _{1})$ and $
V(x;N,M) $ are strictly isospectral, it is natural to wonder whether $
\widetilde{V}^{(n^{2})}(x;N,M,\lambda _{1})$ shares the same
invariance property as $V(x;N,M) $.

Consider the SUSY partner of $\widetilde{V}^{( n^{2})
}(x;N,M,\lambda _{1})$. It is given by
\begin{gather*}
  \widetilde{V}_{\rm SUSY}^{(n^{2})}(x;N,M,\lambda _{1}) =
    \widetilde{V}^{(n^{2})}(x;N,M,\lambda _{1})-2\big( \log
    \widetilde{\psi }_{0}^{(n^{2})}(x;N,M,\lambda _{1})\big)
    ^{\prime \prime } \\
  \qquad {}= V(x;N,M) -2\big[ \log \big( \psi _{n}(x;N,M)\Psi
    _{n}^{(n) }(x;N,M,\lambda _{1})\widetilde{\psi }_{0}^{(
    n^{2}) }(x;N,M,\lambda _{1})\big) \big] ^{\prime \prime },
\end{gather*}
where, for $n\geq 1$,
\begin{gather*}
  \widetilde{\psi }_{0}^{(n^{2})}(x;N,M,\lambda _{1})=E_{n}(
  N,M) \psi _{0}(x;N,M) -\frac{W(\psi _{n},\psi _{0}\,|\, x)}{
  \Psi _{n}^{(n) }(x;N,M,\lambda _{1})}
\end{gather*}
and, for $n=0$,
\begin{gather*}
  \widetilde{\psi }_{0}^{(0^{2})}(x;N,M,\lambda _{1})=\frac{1}{%
  \Psi _{0}^{(0) }(x;N,M,\lambda _{1})}.
\end{gather*}

Consequently, in the $n=0$ case, we have
\begin{gather*}
  \widetilde{V}_{\rm SUSY}^{(0^{2})}(x;N,M,\lambda _{1})  = V (
    x;N,M ) \\
\hphantom{\widetilde{V}_{\rm SUSY}^{(0^{2})}(x;N,M,\lambda _{1})  =}{}
-2\big[ \log \big( \psi _{0}(x;N,M) \Psi _{0}^{( 0) }(x;N,M,\lambda _{1})\widetilde{\psi }_{0}^{( 0^{2})
    }(x;N,M,\lambda _{1})\big) \big] ^{\prime \prime } \\
\hphantom{\widetilde{V}_{\rm SUSY}^{(0^{2})}(x;N,M,\lambda _{1})  =}{}
  = V(x;N,M) -2[ \log \psi _{0}(x;N,M)] ^{\prime\prime },
\end{gather*}
that is
\begin{gather*}
  \widetilde{V}_{\rm SUSY}^{(0^{2})}(x;N,M,\lambda _{1})=V^{(
  0) }(x;N,M) =V(x;N+1,M+1) +E_{1}(N,M) .
\end{gather*}

Furthermore, in the $n\geq 1$ case, we get
\begin{gather*}
 \widetilde{V}_{\rm SUSY}^{(n^{2})}(x;N,M,\lambda _{1})
= V(x;N,M) -2 [ \log \psi _{0}(x;N,M)] ^{\prime \prime } \\
\qquad \quad{}-2\left\{ \log \left[ \left( E_{n}(N,M) \Psi _{n}^{(
    n) }(x;N,M,\lambda _{1})+\frac{W(\psi _{0},\psi _{n}\,|\, x)}{\psi
    _{0}(x;N,M) }\right) \psi _{n}(x;N,M) \right]
    \right\} ^{\prime \prime } \\
\qquad {}= V(x;N+1,M+1) +E_{1}(N,M) \\
 \qquad \quad {}-2\biggl\{ \log \left[ \left( E_{n}(N,M) \Psi _{n}^{(
    n) }(x;N,M,\lambda _{1})-\frac{E_{n}(N,M) }{4n}\psi
    _{n-1}(x;N+1,M+1) \right)\right] \\
\qquad \quad{}+ \log\left[ \psi _{n}(x;N,M) \right]\biggr\} ^{\prime \prime }.
\end{gather*}

More precisely
\begin{gather*}
   \widetilde{V}_{\rm SUSY}^{(n^{2})}(x;N,M,\lambda _{1})
 = V(x;N+1,M+1) -2 [ \log \psi _{n-1}(x;N+1,M+1)  ]
    ^{\prime \prime }+E_{1}(N,M) \\
  \qquad \quad {}-2\left( \log \frac{\lambda _{1}+Q_{n}^{(N,M) } (
     z  ) -\psi _{n}(x;N,M) \psi _{n-1} (
     x;N+1,M+1 ) /4n}{\psi _{n-1}(x;N+1,M+1) }\right) ^{\prime\prime } \\
  \qquad {}= V^{( n-1) }(x;N+1,M+1) +E_{1}(N,M) \\
\qquad \quad{}-2\left( \log \frac{\lambda _{1}+Q_{n}^{(N,M) } (
    z  ) -\psi _{n}(x;N,M) \psi _{n-1} (
    x;N+1,M+1 ) /4n}{\psi _{n-1}(x;N+1,M+1) }\right) ^{\prime\prime }.
\end{gather*}

If, for an appropriate constant $C$, the following condition
\begin{gather}
  \lambda _{1}+Q_{n}^{(N,M) }(z) -\frac{\psi
  _{n}(x;N,M) \psi _{n-1}(x;N+1,M+1) }{4n}=C\big(
  \lambda _{1}^{\prime }+Q_{n-1}^{(N+1,M+1) }(z)
  \big)   \label{cond}
\end{gather}
is satisf\/ied, then
\begin{gather*}
   \frac{\lambda _{1}+Q_{n}^{(N,M) }(z) -\psi
    _{n}(x;N,M) \psi _{n-1}(x;N+1,M+1) /4n}{\psi
    _{n-1}(x;N+1,M+1) }
 =C\Psi _{n-1}^{(n-1)}(x;N+1,M+1,\lambda _{1}^{\prime })
\end{gather*}
and we obtain an enlarged shape invariance property
\begin{gather}
  \widetilde{V}_{\rm SUSY}^{(n^{2})}(x;N,M,\lambda _{1})=\widetilde{V}
  ^{( (n-1) ^{2}) }(x;N+1,M+1,\lambda _{1}^{\prime
  })+E_{1}(N,M) . \label{ESI}
\end{gather}

The preceding condition (\ref{cond}) can be rewritten as
\begin{gather}
  A(z) -B(z)=C\lambda _{1}^{\prime }-\lambda _{1},\qquad \forall\,   z\in {}]{-}1,1[  \label{condb}
\end{gather}
with
\begin{gather*}
  A(z)  = Q_{n}^{(N,M) }(z) -CQ_{n-1}^{(
    N+1,M+1) }(z) \\
 \hphantom{A(z)}{} = \frac{1}{2}\int_{-1}^{z}d\zeta \Big[ C(1-\zeta)
    ^{N+1}(1+\zeta) ^{M+1} \big( P_{n-1}^{(N+1,M+1)
    }(\zeta) \big) ^{2}\\
 \hphantom{A(z)=}{}  -(1-\zeta) ^{N} ( 1+\zeta
   ) ^{M}\big( P_{n}^{(N,M) }(\zeta) \big)^{2}\Big]
\end{gather*}
and
\begin{gather*}
  B(z)  = \psi _{n}(x;N,M) \psi _{n-1}(x;N+1,M+1) /4n \\
\hphantom{B(z)}{} =\frac{(1-z) ^{N+1}(1+z) ^{M+1}}{4n}P_{n}^{( N,M) }(z) P_{n-1}^{(N+1,M+1) }(z) .
\end{gather*}

Equation (\ref{condb}) is equivalent to
\begin{gather}
  \frac{d}{dz}A(z) =\frac{d}{dz}B(z),  \label{condb'}
\end{gather}
where $\frac{d}{dz}A(z) $ is given by
\begin{gather*}
  \frac{2}{(1-z) ^{N}(1+z) ^{M}}\frac{d}{dz}A(
  z) =C\big( 1-z^{2}\big) \big( P_{n-1}^{(N+1,M+1)
  }(z) \big) ^{2}-\big( P_{n}^{(N,M) } (
  z ) \big) ^{2}.  
\end{gather*}
\par

As for $\frac{d}{dz}B(z)$, using the derivation formula \cite{magnus, szego}
\begin{gather}
  \frac{d}{dz}P_{n}^{(N,M) }(z) =\frac{N+M+n+1}{2}  P_{n-1}^{(N+1,M+1) }(z) ,  \label{deriv}
\end{gather}
it can be expressed as
\begin{gather*}
    \frac{4n}{(1-z) ^{N}(1+z) ^{M}}\frac{d}{dz}B (
    z ) = \big( 1-z^{2}\big) P_{n}^{(N,M) }(z)
    \frac{d}{dz}P_{n-1}^{(N+1,M+1) }(z) \\
   \qquad {}+\big( 1-z^{2}\big) \frac{N+M+n+1}{2}\big( P_{n-1}^{(
    N+1,M+1) }(z) \big) ^{2} \\
 \qquad {}+ [ (M-N)-z ( N+M+2 )  ] P_{n}^{(N,M)
    }(z) P_{n-1}^{(N+1,M+1) }(z) ,
\end{gather*}
that is
\begin{gather*}
   \frac{4n}{(1-z) ^{N}(1+z) ^{M}}\frac{d}{dz}B (
    z ) = \big( 1-z^{2}\big) \frac{N+M+n+1}{2}\big( P_{n-1}^{ (
    N+1,M+1 ) }(z) \big) ^{2} \\
  \qquad {}+\big( 1-z^{2}\big) P_{n}^{(N,M) }(z) \frac{d}{
    dz}P_{n-1}^{(N+1,M+1) }(z) \\
  \qquad {}+ [ (M-N)-z ( N+M+2 )  ] P_{n}^{(N,M)
    }(z) P_{n-1}^{(N+1,M+1) }(z) .
\end{gather*}

But the dif\/ferential equation satisf\/ied by the Jacobi polynomials is \cite{magnus, szego}
\begin{gather*}
 \big( 1-z^{2}\big) \frac{d^{2}}{dz^{2}}P_{n}^{(N,M) } (
    z ) + [ (M-N)-z ( N+M+2 )  ] \frac{d}{dz}
    P_{n}^{(N,M) }(z)  \\
    \qquad{} = -n ( N+M+n+1 ) P_{n}^{(N,M) }(z) ,
\end{gather*}
which, combined with equation~(\ref{deriv}), gives
\begin{gather*}
\big( 1-z^{2}\big) \frac{d}{dz}P_{n-1}^{(N+1,M+1) } (
    z ) + [ (M-N)-z ( N+M+2 )  ] P_{n-1}^{(
    N+1,M+1) }(z)  =-2nP_{n}^{(N,M) }(z) .
\end{gather*}

Consequently
\begin{gather*}
  \frac{4n}{(1-z) ^{N}(1+z) ^{M}}\frac{d}{dz}B(z) \\
  \qquad{}
 =-2n\big( P_{n}^{(N,M) }(z) \big)
    ^{2}+\big( 1-z^{2}\big) \frac{N+M+n+1}{2}\big( P_{n-1}^{ (
    N+1,M+1) }(z) \big) ^{2},
\end{gather*}
or
\begin{gather*}
   \frac{2}{(1-z) ^{N}(1+z) ^{M}}\frac{d}{dz}B(z) \\
  \qquad {} =\big( 1-z^{2}\big) \frac{N+M+n+1}{4n}\big( P_{n-1}^{(
    N+1,M+1) }(z) \big) ^{2}-\big( P_{n}^{(N,M)
    }(z) \big) ^{2}.
\end{gather*}

To satisfy the equality (\ref{condb'}), we then must choose
\begin{gather*}
  C=\frac{N+M+n+1}{4n}.
\end{gather*}
In this case, equation~(\ref{condb}) simply becomes
\begin{gather*}
  A(z) -B(z)=\frac{N+M+n+1}{4n}\lambda _{1}^{\prime }-\lambda _{1}.
\end{gather*}

In the limit $z \rightarrow 1^-$, we obviously get
\begin{gather*}
  B(1)=0,
\end{gather*}
and (see equations~(\ref{Q1}) and (\ref{recQ}))
\begin{gather*}
  A(1)=Q_{n}^{(N,M) }(1) -\frac{N+M+n+1}{4n}
  Q_{n-1}^{(N+1,M+1) }(1) =0,
\end{gather*}
which implies
\begin{gather}
  \lambda _{1}^{\prime }=\frac{4n}{N+M+n+1}\lambda _{1}.  \label{tranflambda}
\end{gather}

We conclude that for $n\ge 1$, the two-step conf\/luent rational extensions of the TDPT potential satisfy the enlarged shape invariance property~(\ref{ESI}) with $\lambda'_1$ given in equation~(\ref{tranflambda}). Note that the latter relation ensures that the domain of~$\lambda _{1}$ values for which $\widetilde{V}^{( n^{2})
}(x;N,M,\lambda _{1})$ is regular corresponds exactly to the domain of $\lambda' _{1}$ values for which $\widetilde{V}^{((n-1)^{2}) }(x;N+1,M+1,\lambda _{1}^{\prime })$ is also regular (see equations~(\ref{recQ}) and~(\ref{reglambda})).

In the $n=N=M=1$ case, such a property can be directly verif\/ied as follows. We get
\begin{gather*}
  P_{0}^{(2,2) }(z) =1,\qquad P_{1}^{(1,1)
  }(z) =2z,
\end{gather*}
$C=1$, $\lambda _{1}^{\prime }=\lambda _{1}$, as well as
\begin{gather*}
  A(z)  = Q_{1}^{(1,1) }(z) -Q_{0}^{(
    2,2) }(z) =\frac{1}{2}\int_{-1}^{z} d\zeta \big( \big(
    1-\zeta ^{2}\big) ^{2}-4\zeta ^{2}\big( 1-\zeta ^{2}\big) \big) \\
\hphantom{A(z)}{}
=\frac{1}{2}\int_{-1}^{z}d\zeta \big( 1-6\zeta ^{2}+5\zeta ^{4}\big)
 =\frac{1}{2}z\big( 1-z^{2}\big) ^{2}
\end{gather*}
and
\begin{gather*}
  B(z)=\frac{\big( 1-z^{2}\big) ^{2}}{4}2z.
\end{gather*}
Hence the identity
\begin{gather*}
  A(z) =B(z),\qquad \forall\, z\in {}]{-}1,1 [,
\end{gather*}
is satisf\/ied, which implies that
\begin{gather*}
  \widetilde{V}_{\rm SUSY}^{( 1^{2}) }(x;1,1,\lambda _{1})=\widetilde{V}
  ^{( (0) ^{2}) }(x;2,2,\lambda _{1})+E_{1}(1,1) .
\end{gather*}

\section{Two-step conf\/luent rational extensions\\ of the isotonic potential}
\label{section5}

\subsection{General scheme}

The isotonic oscillator potential (with zero ground-state energy $E_{0}=0$) is
def\/ined on the positive half-line $] 0,+\infty [ $ by
\begin{gather*}
  V(x;\omega,\alpha) =\frac{\omega ^{2}}{4}x^{2}+\frac{(
  \alpha +1/2) (\alpha -1/2)}{x^{2}}-\omega (\alpha+1)
  ,\qquad  \vert \alpha  \vert >1/2. 
\end{gather*}

If we add Dirichlet boundary conditions at zero and inf\/inity and if we
suppose $\alpha >1/2$, it has the following spectrum ($z=\omega x^{2}/2$)
\begin{gather*}
    E_{n}( \omega ) =2n\omega, \qquad
    \psi _{n}(x;\omega,\alpha) =z^{( \alpha +1/2)
      /2}e^{-z/2}L_{n}^{\alpha }(z),
\qquad n\geq 0,  
\end{gather*}
where
\begin{gather}
  L_{n}^{\alpha }(z) =\sum_{k=0}^{n}\frac{(-1) ^{k}}{k!
  }\frac{( n+\alpha ) \cdots  ( k+1+\alpha  ) }{(  n-k) !}z^{k}  \label{lag}
\end{gather}
is the usual Laguerre polynomial \cite{magnus,szego}.

In the following, in order to get rational extensions, we consider the case where $\alpha $ is an integer: $\alpha =N \ge 1$.

If we choose as initial seed function an (unnormalized) eigenstate $\psi
_{n}(x;\omega,N) $ of $V(x;\omega,N) $, by taking $
x_{0}=0$ we get
\begin{gather*}
  \int_{0}^{x}dt\psi _{n}^{2}(t;\omega ,N)  =  \Vert \psi _{n} \Vert
    ^{2}-\int_{x}^{+\infty }dt\psi _{n}^{2}(t;\omega ,N) \\
\hphantom{\int_{0}^{x}dt\psi _{n}^{2}(t;\omega ,N)}{}
=  \Vert \psi _{n} \Vert ^{2}-\frac{1}{\sqrt{2\omega }}
    \int_{z}^{+\infty }d\zeta \zeta ^{N}\big( L_{n}^{N}(\zeta)
    \big) ^{2}\exp ( -\zeta  ) .
\end{gather*}

But
\begin{gather}
  F_{n}^{N}(z)=-\frac{1}{\sqrt{2\omega }}\int_{z}^{+\infty }d\zeta \zeta
  ^{N}\big( L_{n}^{N}(\zeta) \big) ^{2}\exp  ( -\zeta
  ) =\frac{1}{\sqrt{2\omega }}\exp (-z) Q_{n}^{N}(z),
  \label{F}
\end{gather}
where
\begin{gather}
  Q_{n}^{N}(z)=-\sum_{j=0}^{N+2n}\frac{d^{j}}{dz^{j}}\big( z^{N}\big(
  L_{n}^{N}(z) \big) ^{2}\big)  \label{Qder}
\end{gather}
is a polynomial of degree $N+2n$.

In contrast with what happened in the TDPT case, to obtain a rational extension
we need now to eliminate the exponential factor, which necessitates f\/ixing $
\lambda _{1}=- \Vert \psi _{n} \Vert ^{2}$. Equation~(\ref{2stepgenform})
then becomes
\begin{gather*}
  \widetilde{V}^{(n^{2})}\big(x;\omega ,N,- \Vert \psi
    _{n} \Vert ^{2}\big)  = V(x;\omega,N) -2\big[ \log \big(
    F_{n}^{N}(z)\big) \big] ^{\prime \prime } \\
\hphantom{\widetilde{V}^{(n^{2})}\big(x;\omega ,N,- \Vert \psi
    _{n} \Vert ^{2}\big)}{}
= V(x;\omega,N) -4\omega z^{1/2}\frac{d}{dz}\left( \frac{
    z^{N+1/2}\big( L_{n}^{N}(z) \big) ^{2}}{Q_{n}^{N}(z)}\right) ,
\end{gather*}
where $\widetilde{V}^{(n^{2})}(x;\omega ,N,- \Vert \psi
_{n} \Vert ^{2})$ constitutes a rational extension of $V(x;\omega,N)$. Since $F_{n}^{N}(z)$ is a negative, strictly increasing function of~$z$, such that
\begin{gather*}
 \lim_{x\rightarrow +\infty} F_{n}^{N}(z)=0,
\end{gather*}
$Q_{n}^{N}(z)$ keeps a constant strictly negative sign, i.e., $Q_{n}^{N}(z)$
has no zero on the positive half-line. This shows that $\widetilde{V}^{(
n^{2}) }(x;\omega ,N,-\Vert \psi _{n}\Vert ^{2})$ is also
regular.

Let us note too that \cite{magnus,szego}
\begin{gather}
  F_{n}^{N}(0)=-\frac{1}{\sqrt{2\omega }}\int_{0}^{+\infty }d\zeta \zeta
  ^{N}\big( L_{n}^{N}(\zeta) \big) ^{2}\exp  ( -\zeta
 ) =-\frac{1}{\sqrt{2\omega }}\frac{(n+N) !}{n!}<0,
  \label{F0}
\end{gather}
or
\begin{gather}
  Q_{n}^{N}(0)=-\frac{(n+N) !}{n!}.  \label{Q0}
\end{gather}

The eigenfunctions of $\widetilde{V}^{(
n^{2}) }(x;\omega ,N,-\Vert \psi _{n}\Vert ^{2})$ are given by (see equation~(\ref{2stepgeneig2}))
\begin{gather*}
    \widetilde{\psi }_{k}^{(n^{2})}\big(x;\omega ,N,- \Vert \psi
    _{n} \Vert ^{2}\big)
 = [ E_{n}(\omega )-E_{k}(\omega ) ] \psi
    _{k}(x;\omega,N) -\frac{W(\psi _{n},\psi _{k}\,|\, x)}{\Psi
    _{n}^{(n) }(x;\omega ,N,- \Vert \psi _{n} \Vert ^{2})},\!\!\!
    \qquad k\neq n,
\end{gather*}
where
\begin{gather}
  \Psi _{n}^{(n) }\big(x;\omega ,N,- \Vert \psi _{n} \Vert
    ^{2})  = \frac{\int_{0}^{x}dt\psi _{n}^{2}(t;\omega ,N)- \Vert \psi
    _{n} \Vert ^{2}}{\psi _{n}(x;\omega ,N)} \nonumber \\
\hphantom{\Psi _{n}^{(n) }\big(x;\omega ,N,- \Vert \psi _{n} \Vert
    ^{2})}{}
=\frac{F_{n}^{N}(z)}{\psi _{n}(x;\omega ,N)}=\frac{1}{\sqrt{2\omega }}
    z^{-(N+1/2) /2}\exp  ( -z/2 ) \frac{Q_{n}^{N}(z)}{
    \mathit{L}_{n}^{N}(z) } \label{Psi}
\end{gather}
and
\begin{gather*}
  W( \psi _{n},\psi _{k}\,|\, x) =\sqrt{2\omega }z^{N+1}\exp (
  -z) L_{n,k}^{N}(z) ,
\end{gather*}
with
\begin{gather*}
  L_{n,k}^{N}(z) =L_{n-1}^{N+1}(z) L_{k}^{N}(
  z) -L_{n}^{N}(z) L_{k-1}^{N+1}(z) .
\end{gather*}

Consequently
\begin{gather*}
  \widetilde{\psi }_{k}^{(n^{2})}\big(x;\omega ,N,- \Vert \psi
    _{n} \Vert ^{2}\big) =  [ E_{n}(\omega )-E_{k}(\omega ) ] \psi
    _{k}(x;\omega,N) \\
\hphantom{\widetilde{\psi }_{k}^{(n^{2})}\big(x;\omega ,N,- \Vert \psi
    _{n} \Vert ^{2}\big) =} {}
    -2\omega z^{( 3N+5/2) /2}\exp  ( -z/2 ) \frac{L_{n,k}^{N}(z) L_{n}^{N}(z) }{Q_{n}^{N}(z)},
\end{gather*}
that is,
\begin{gather*}
  \widetilde{\psi }_{k}^{(n^{2})}\big(x;\omega ,N,- \Vert \psi
  _{n} \Vert ^{2}\big)=\frac{2\omega z^{(N+1/2) /2}\exp  (
  -z/2 ) }{Q_{n}^{N}(z)}\widetilde{L}_{k,N}^{(n^{2})} (
  z ) ,
\end{gather*}
where
\begin{gather*}
  \widetilde{L}_{k,N}^{(n^{2})}(z) =(n-k)
  L_{k}^{N}(z) Q_{n}^{N}(z)-z^{N+1}L_{n,k}^{N}(z)
  L_{n}^{N}(z) .
\end{gather*}

$\widetilde{\psi }_{k}^{( n^{2}) }(x;\omega ,N,-\Vert \psi
_{n}\Vert ^{2})$ decreases exponentially to zero at inf\/inity and also tends
to zero at the origin (see equation~(\ref{Q0})). It is therefore an admissible eigenstate of $\widetilde{V}^{( n^{2}) }(x;\omega ,N,-\Vert \psi _{n}\Vert ^{2})$
for every $k\neq n$.

Note that in this case
\begin{gather}
  \widetilde{\psi }_{n}^{(n^{2})}\big(x;\omega ,N,-\Vert \psi
  _{n} \Vert ^{2}\big)=1/\Psi _{n}^{(n) }\big(x;\omega ,N,- \Vert
  \psi _{n} \Vert ^{2}\big)\propto z^{(N+1/2) /2}\exp  (
  z/2 ) \frac{\mathit{L}_{n}^{N}(z) }{Q_{n}^{N}(z)}
  \label{no-state}
\end{gather}
is not an eigenstate since it is not normalizable. This implies that $
\widetilde{V}^{(n^{2})}(x;\omega ,N,-\Vert \psi
_{n}\Vert ^{2})$ and $V(x;\omega ,N)$ are only quasi-isospectral, the
two-step conf\/luent DBT $A(W_{\nu }^{(\nu) })\circ A(w_{\nu })$
being now state-deleting.

The orthogonality conditions between eigenstates imply that the set of $\widetilde{L
}_{k,N}^{(n^{2})}(z)$ with $k\neq n,$ constitute a
family of orthogonal polynomials, indexed by $k\in \mathbb{N}$, on the positive half-line with respect to the mesure
\begin{gather*}
  \mu _{N}^{(n^{2})}(z)=\frac{z^{N}\exp (-z) }{\big(
  Q_{n}^{N}(z)\big) ^{2}}.
\end{gather*}

\subsection{Examples}

\subsubsection[The $n=0$ case]{The $\boldsymbol{n=0}$ case}

The $n=0$ case gives no new extended potentials with respect to previously known
results. Indeed, the corresponding conf\/luent two-step DBT is obtained by
applying successively the one-step DBT $A(w_{0})$ and $A(W_{0}^{(
0) })$. The f\/irst one is the usual SUSY partnership and, due to the
shape invariance property of the isotonic potential, we have
\begin{gather*}
  V^{(0) }(x;\omega ,N)=V(x;\omega ,N+1)+2\omega ,
\end{gather*}
where $V^{(0) }(x;\omega ,N)$ admits the following eigenstates
\begin{gather*}
  \psi _{k}^{(0) }(x;\omega ,N)\sim \psi _{k-1}(x;w,N+1),
\end{gather*}
for the respective energies $E_{k}( \omega)$, $k\geq 1$.

The second DBT $A\bigl(W_{0}^{(0) }\bigr)$ is associated to the seed
function $\Psi _{0}^{(0) }(x;\omega,N,-\Vert \psi
_{0}\Vert ^{2})$, which is a formal
eigenfunction of $V^{(0) }(x;\omega ,N)$ for the eigenvalue $E_{0}=0$ and which is then in the disconjugacy sector of $V^{(0)
}(x;\omega ,N)$. But
\begin{gather*}
  \Psi _{0}^{(0) }(x;\omega ,N,\lambda _{1})=\psi _{0}^{(
  0) }(x;\omega ,N)\left( \lambda _{1}+\int_{x_{0}}^{x}dt\frac{1}{\big(
  \psi _{0}^{(0) }(t;\omega ,N)\big) ^{2}}\right) ,
\end{gather*}
where
\begin{gather*}
  \psi _{0}^{(0) }(x;\omega ,N)\sim 1/\psi _{0}(x;\omega
  ,N)=z^{-(N+1/2)/2}e^{+z/2}.
\end{gather*}

By taking $x_{0}=0$ and $\lambda _{1}=-\left\Vert \psi _{0}\right\Vert ^{2}$, we obtain
\begin{gather}
  \Psi _{0}^{(0) }\big(x;\omega ,N,- \Vert \psi _{0} \Vert
  ^{2}\big)=\frac{1}{\sqrt{2\omega }}z^{-(N+1/2)/2}\exp  ( -z/2 )
  Q_{0}^{N}(z),  \label{seed02}
\end{gather}
where
\begin{gather*}
  Q_{0}^{N}(z)=-\sum_{j=0}^{N}\frac{d^{j}}{dz^{j}}\big( z^{N}\big)
  =-N!\sum_{l=0}^{N}\frac{z^{l}}{l!},
\end{gather*}
which gives
\begin{gather}
  \Psi _{0}^{(0) }\big(x;\omega ,N,- \Vert \psi _{0} \Vert
  ^{2}\big)=-\frac{N!}{\sqrt{2\omega }}\exp  ( -z/2 )
  z^{-(N+1/2)/2}\sum_{l=0}^{N}\frac{z^{l}}{l!}.  \label{seed0}
\end{gather}

Equation~(\ref{seed0}) shows that $\Psi _{0}^{(0) }(x;\omega ,N,- \Vert \psi _{0} \Vert
^{2})$ is a formal eigenfunction of $V^{(0) }(x;\omega
,N)=V(x;\omega ,N+1)+2\omega $ that tends to zero at inf\/inity and diverges
at the origin. Such an eigenfunction is unique (up to a constant multiplicative factor) at
a given eigenvalue (here $E_{0}=0$) \cite{berezin,hartman}.

Consequently $\Psi _{0}^{(0) }(x;\omega ,N,-\Vert \psi
_{0}\Vert ^{2})$ coincides necessarily (up to a constant multiplicative factor) with
a type II seed function~\cite{GGM} of $V(x;\omega ,N+1)+2\omega $, which is
a formal eigenfunction for the eigenvalue $E_{-1}+2\omega $, namely
\begin{gather*}
  \Psi _{0}^{(0) }\big(x;\omega ,N,-\Vert \psi _{0}\Vert
  ^{2}\big)\sim \phi _{N,-}(x;\omega ,N+1)=z^{-(N+1/2) /2}e^{-z/2}  L_{N}^{-N-1}(z) ,  \label{type2}
\end{gather*}
where $\phi _{n,-}$ is def\/ined by
\begin{gather*}
  \phi _{n,-}(x;\omega ,\alpha )=\psi _{n}(x;\omega ,-\alpha ).
\end{gather*}

Note that this result can be recovered directly by using the identity (see
equation~(\ref{lag}))
\begin{gather*}
  L_{N}^{-N-1}(z) =(-1) ^{N}\sum_{l=0}^{N}\frac{z^{l}}{l!},
\end{gather*}
in equation~(\ref{seed0}).

Applying then the DBT $A\big(W_{0}^{(0) }\big)=A(v_{N,-}(x;\omega ,N+1))$
($v_{N,-}=-\phi _{N,-}^{\prime }/\phi _{N,-}$) to $V^{(0)
}(x;$ $\omega ,N)=V(x;\omega ,N+1)+2\omega $, we arrive at a two-step conf\/luent
extended potential
\begin{gather*}
  \widetilde{V}^{(0^{2})}\big(x;\omega ,N,-\Vert \psi
  _{0}\Vert ^{2}\big)=V^{( N,-) }(x;\omega ,N+1)+2\omega ,
\end{gather*}
which coincides (up to a constant shift) with the usual extensions obtained
with type~II seed functions~\cite{GGM}.

Explicitly, we get
\begin{gather*}
  \widetilde{V}^{(0^{2})}\big(x;\omega ,N,-\Vert \psi
  _{0}\Vert ^{2}\big)=V(x;\omega,N) +\frac{4\omega }{N!}z^{1/2}
  \frac{d}{dz}\left[ z^{N+1/2}\Big/\left( \sum\limits_{l=0}^{N}z^{l}/l!\right)\right]
\end{gather*}
and
\begin{gather*}
  \widetilde{\psi }_{k}^{(0^{2})}\big(x;\omega ,N,- \Vert \psi
  _{0} \Vert ^{2}\big)\sim \widetilde{L}_{k,N}^{(0^{2})} (
  z ) \frac{z^{(N+1/2) /2}\exp  ( -z/2 ) }{
  \sum\limits_{l=0}^{N}z^{l}/l!},\qquad  k\geq 1.
\end{gather*}

In the $N=1$ case, this gives
\begin{gather*}
  \widetilde{V}^{(0^{2})}\big(x;\omega ,1,- \Vert \psi
    _{0} \Vert ^{2}\big)  =\frac{\omega ^{2}}{4}x^{2}+\frac{3}{4x^{2}}+\frac{
    4\omega }{\omega x^{2}+2}-\frac{16\omega }{( \omega x^{2}+2) ^{2}}
  = V^{ ( 1,- ) }(x;\omega ,2)+2\omega
\end{gather*}
and
\begin{gather*}
  \widetilde{\psi }_{k}^{(0^{2})}\big(x;\omega ,1,-\Vert \psi
  _{0}\Vert ^{2}\big)\sim \frac{2\omega z^{3/4}\exp  ( -z/2 ) }{z+1}
  \widetilde{L}_{k,1}^{(0^{2})}(z) ,\qquad k\geq 1,
\end{gather*}
with
\begin{gather*}
  \widetilde{L}_{k,1}^{(0^{2})}(z) =k\mathit{L}
  _{k}^{1}(z) (z+1) +z^{2}\mathit{L}_{k-1}^{2}(
  z) .
\end{gather*}

\subsubsection[The $n=1$ case]{The $\boldsymbol{n=1}$ case}

We have in this case
\begin{gather*}
  Q_{1}^{N}(z)=-z^{N+2}+Nz^{N+1}-(N+1)!\sum_{j=0}^{N}\frac{z^{j}}{j!}
\end{gather*}
and
\begin{gather*}
   \widetilde{V}^{( 1^{2}) }\big(x;\omega ,N,- \Vert \psi
    _{1} \Vert ^{2}\big) = V(x;\omega,N)
 -4\omega \frac{z^{N} ( N+1-z ) }{Q_{1}^{N}(z)}\big[ \big(
    N+\tfrac{1}{2}\big)  ( N+1 ) -\big( N+\tfrac{5}{2}\big) z\big]  \\
  \qquad {}-4\omega \frac{z^{N+1} ( N+1-z ) ^{2}}{\big[ Q_{1}^{N}(z)\big]
    ^{2}} \left(  ( N+2 ) z^{N+1}-N ( N+1 )
    z^{N}+(N+1)!\sum_{j=0}^{N-1}\frac{z^{j}}{j!}\right) .
\end{gather*}

The eigenstates of $\widetilde{V}^{( 1^{2}) }$ are ($k\neq 1$)
\begin{gather*}
  \widetilde{\psi }_{k}^{( 1^{2}) }\big(x;\omega ,N,- \Vert \psi
    _{1} \Vert ^{2}\big)  =  ( 1-k ) \psi _{k}(x;\omega,N)
 -z^{( 3N+5/2) /2}\exp ( -z/2) \frac{
L_{1,k}^{N}(z) L_{1}^{N}(z) }{Q_{1}^{N}(z)},
\end{gather*}
that is,
\begin{gather*}
  \widetilde{\psi }_{k}^{( 1^{2}) }\big(x;\omega ,N,- \Vert \psi
  _{1} \Vert ^{2}\big)=\frac{2\omega z^{(N+1/2) /2}\exp (
  -z/2) }{Q_{1}^{N}(z)}\widetilde{L}_{k,N}^{( 1^{2}) }(
  z) ,
\end{gather*}
where
\begin{gather*}
  \widetilde{L}_{k,N}^{( 1^{2}) }(z) = z^{N+1}(
    N+1-z) ^{2}L_{k-1}^{N+1}(z) \\
\hphantom{\widetilde{L}_{k,N}^{( 1^{2}) }(z) =}{}
+\left( kz^{N+2}- ( Nk+1 ) z^{N+1}+ ( k-1 )
(N+1)!\sum_{j=0}^{N}\frac{z^{j}}{j!}\right) L_{k}^{N}(z) .
\end{gather*}

In particular, for $N=1$
\begin{gather*}
  Q_{1}^{1}(z)=-z^{3}+z^{2}-2z-2
\end{gather*}
and
\begin{gather*}
  \widetilde{V}^{( 1^{2}) }\big(x;\omega ,1,-\Vert \psi
    _{1} \Vert ^{2}\big)  =V ( x;\omega ,1 ) +6\omega\frac{z^{2}-4}{
    z^{3}-z^{2}+2z+2}
    -20\omega\frac{5z^{2}-4z-2}{\big( z^{3}-z^{2}+2z+2\big) ^{2}}
    + 2\omega,
\end{gather*}
or
\begin{gather*}
  \widetilde{V}^{( 1^{2}) }\big(x;\omega ,1,- \Vert \psi
    _{1} \Vert ^{2}\big)  =\frac{\omega ^{2}}{4}x^{2}+\frac{3}{4x^{2}}
    +12\omega \frac{\omega ^{2}x^{4}-16}{\omega ^{3}x^{6}-2\omega
    ^{2}x^{4}+8\omega x^{2}+16} \\
  \hphantom{\widetilde{V}^{( 1^{2}) }\big(x;\omega ,1,- \Vert \psi
    _{1} \Vert ^{2}\big)  =}{}
 -320\omega \frac{5\omega ^{2}x^{4}-8\omega x^{2}-8}{\big( \omega
    ^{3}x^{6}-2\omega ^{2}x^{4}+8\omega x^{2}+16\big) ^{2}}.
\end{gather*}
Moreover ($k\neq 1$)
\begin{gather*}
  \widetilde{\psi }_{k}^{( 1^{2}) }\big(x;\omega ,1,- \Vert \psi
  _{1} \Vert ^{2}\big)=-\frac{2\omega z^{3/4}\exp ( -z/2 ) }{
  z^{3}-z^{2}+2z+2}\widetilde{L}_{k,1}^{( 1^{2}) }(z) ,
\end{gather*}
where
\begin{gather*}
  \widetilde{L}_{k,1}^{( 1^{2}) }(z) =\big[
  kz^{3}-( k+1) z^{2}+2( k-1) z+2( k-1)
  \big] L_{k}^{1}(z) +z^{2}( 2-z)
  ^{2}L_{k-1}^{2}(z) .
\end{gather*}

The polynomial
\begin{gather*}
  \widetilde{L}_{0,1}^{( 1^{2}) }(z) =-\big(
  z^{2}+2z+2\big)
\end{gather*}
is associated to the ground state of $\widetilde{V}^{( 1^{2})
}\big(x;\omega ,1,-\Vert \psi _{1}\Vert ^{2}\big)$ and
\begin{gather*}
  \widetilde{L}_{2,1}^{( 1^{2}) }(z) =-\frac{1}{2}  \big( z^{4}+4z^{2}-12\big) ,
\end{gather*}
which has a single zero at $\sqrt{2}$ on $] 0,+\infty [ $,
corresponds to the f\/irst excited eigenstate.

\subsection{Shape invariance of the two-step conf\/luent rational extensions\\
of the isotonic potential}

$V(x;\omega,N) $ is a translationally shape invariant potential
with a SUSY partner
\begin{gather*}
  V_{\rm SUSY}(x;\omega ,N)=V^{(0) }(x;\omega,N) =V(
x;\omega ,N+1) +E_{1}( \omega )
\end{gather*}
and (the coef\/f\/icient is readily established by a direct calculation)
\begin{gather}
  \psi _{n}^{(0) }(x;\omega,N) =\frac{W(\psi
  _{0},\psi _{n}\,|\, x)}{\psi _{0}(x;\omega,N) }=-\sqrt{2\omega }
  \psi _{n-1}( x;\omega ,N+1) .  \label{wron}
\end{gather}

Considering the SUSY partner of $\widetilde{V}^{(n^{2})}\big(x;\omega
,N,-\left\Vert \psi _{n}\right\Vert ^{2}\big)$, let us deal separately with the $n\ge 1$ and $n=0$ cases.
For $n\geq 1$, this partner is given by
\begin{gather}
  \widetilde{V}_{\rm SUSY}^{(n^{2})}\big(x;\omega ,N,- \Vert \psi
    _{n} \Vert ^{2}\big) = \widetilde{V}^{(n^{2})}\big(x;\omega
    ,N,- \Vert \psi _{n} \Vert ^{2}\big)-2\big( \log \widetilde{\psi }
    _{0}^{(n^{2})}\big(x;\omega ,N,- \Vert \psi _{n} \Vert
    ^{2}\big)\big) ^{\prime \prime } \!\!\!\!\label{IP-partner} \\
 \qquad {}= V(x;\omega,N)
  -2\big[ \log \big( \psi _{n}(x;\omega ,N)\Psi _{n}^{(n)
    }(x;\omega ,N,- \Vert \psi _{n} \Vert ^{2})\widetilde{\psi }
    _{0}^{(n^{2})}(x;\omega ,N,- \Vert \psi _{n} \Vert
    ^{2})\big) \big] ^{\prime \prime },\nonumber
\end{gather}
where
\begin{gather}
  \widetilde{\psi }_{0}^{(n^{2})}\big(x;\omega ,N,-\Vert \psi
  _{n}\Vert ^{2}\big)=E_{n}(\omega) \psi _{0}( x;\omega
  ,N) -\frac{W(\psi _{n},\psi _{0}\,|\, x)}{\Psi _{n}^{(n)
  }\big(x;\omega ,N,- \Vert \psi _{n} \Vert ^{2}\big)}. \label{IP-gs}
\end{gather}

Inserting (\ref{IP-gs}) into (\ref{IP-partner}), we get
\begin{gather*}
    \widetilde{V}_{\rm SUSY}^{(n^{2})}\big(x;\omega ,N,- \Vert \psi
    _{n} \Vert ^{2}\big)
 = V(x;\omega,N) -2 ( \log \psi
    _{0}(x;\omega ,N) ) ^{\prime \prime } \\
 \qquad\quad {}-2\left\{\log \left[ \left( E_{n}(\omega) \Psi _{n}^{(
    n) }\big(x;\omega ,N,-\Vert \psi _{n}\Vert ^{2}\big)+\frac{W(\psi
    _{0},\psi _{n}\,|\, x)}{\psi _{0}(x;\omega,N) }\right) \psi
    _{n}(x;\omega,N)  \right]\right\} ^{\prime \prime } \\
 \qquad {}= V ( x;\omega ,N+1 ) +E_{1}(\omega) \\
 \qquad \quad {}-2\left\{ \log \left[ \left( 2n\omega \Psi _{n}^{(n) }\big(x;\omega
    ,N,- \Vert \psi _{n} \Vert ^{2}\big)-\sqrt{2\omega }\psi _{n-1} (
    x;\omega ,N+1 ) \right) \psi _{n}(x;\omega,N) \right]
    \right\} ^{\prime \prime }.
\end{gather*}
More precisely, on using (\ref{Psi}),
\begin{gather*}
  \widetilde{V}_{\rm SUSY}^{(n^{2})}\big(x;\omega ,N,- \Vert \psi
    _{n} \Vert ^{2}\big)
 = V ( x;\omega ,N+1 ) -2 ( \log \psi
    _{n-1} ( x;\omega ,N+1 )  ) ^{\prime \prime }+E_{1} (
    \omega  ) \\
 \qquad\quad {}-2\left\{ \log\left[ \left( F_{n}^{N}(z)-\frac{\psi _{n}(x;\omega,N)
    \psi _{n-1} ( x;\omega ,N+1 ) }{\sqrt{2\omega }n}\right) /\psi
    _{n-1} ( x;\omega ,N+1 ) \right] \right\} ^{\prime \prime } \\
 \qquad {}= V^{(n-1) } ( x;\omega ,N+1 ) +E_{1} ( \omega
  )  \\
 \qquad\quad {}-2\left\{ \log\left[ \left( F_{n}^{N}(z)-\frac{\psi _{n}(x;\omega,N)
    \psi _{n-1} ( x;\omega ,N+1 ) }{\sqrt{2\omega }n}\right) /\psi
    _{n-1} ( x;\omega ,N+1 ) \right] \right\} ^{\prime \prime }.
\end{gather*}

If, for an appropriate constant $C$, the following condition
\begin{gather}
  F_{n}^{N}(z)-\frac{\psi _{n}(x;\omega,N) \psi _{n-1} (
  x;\omega ,N+1 ) }{\sqrt{2\omega }n}=CF_{n-1}^{N+1}(z)  \label{cond-IP}
\end{gather}
is satisf\/ied, then
\begin{gather*}
  \left( F_{n}^{N}(z)-\frac{\psi _{n}(x;\omega,N) \psi
    _{n-1}\left( x;\omega ,N+1\right) }{\sqrt{2\omega }n}\right) \Big/\psi
    _{n-1} ( x;\omega ,N+1 ) \\
 \qquad {}=C\Psi _{n-1}^{(n-1)
    }\big(x;\omega ,N+1,- \Vert \psi _{n-1} \Vert ^{2}\big)
\end{gather*}
and we obtain an enlarged shape invariance property
\begin{gather}
  \widetilde{V}_{\rm SUSY}^{(n^{2})}\big(x;\omega ,N,- \Vert \psi
  _{n} \Vert ^{2}\big)=\widetilde{V}^{ ( (n-1) ^{2} )
  }\big(x;\omega ,N+1,- \Vert \psi _{n-1} \Vert ^{2}\big)+E_{1} ( \omega
 ) . \label{ESI-IP}
\end{gather}

The preceding condition (\ref{cond-IP}) can be rewritten as
\begin{gather}
  A(z) -B(z)=0,\qquad \forall \, z\in {}]{-}1,1 [, \label{condb-IP}
\end{gather}
with
\begin{gather*}
  A(z)  = F_{n}^{N}(z)-CF_{n-1}^{N+1}(z)
  = \frac{1}{\sqrt{2\omega }}\int_{z}^{+\infty }d\zeta \Big( C\zeta
    ^{N+1}\big( L_{n-1}^{N+1}(\zeta) \big) ^{2}-\zeta ^{N}\big(
    L_{n}^{N}(\zeta) \big) ^{2}\Big) \exp  ( -\zeta  )
\end{gather*}
and
\begin{gather*}
  B(z) = \frac{\psi _{n}(x;\omega,N) \psi _{n-1}( x;\omega
  ,N+1) }{\sqrt{2\omega }n} = \frac{z^{N+1}e^{-z}L_{n}^{N}(z)
  L_{n-1}^{N+1}(z)
  }{\sqrt{2\omega }n}.
\end{gather*}

Equation (\ref{condb-IP}) is equivalent to the set of conditions
\begin{gather}
  \frac{d}{dz}A(z) =\frac{d}{dz}B(z)  \label{condb'-IP}
\end{gather}
and
\begin{gather}
  A(0) = B(0).  \label{condb"-IP}
\end{gather}

Here $\frac{d}{dz}A(z) $ is given by
\begin{gather*}
  \sqrt{2\omega }\frac{d}{dz}A(z) =\big( \big( L_{n}^{N} (
  z ) \big) ^{2}-Cz\big( L_{n-1}^{N+1}(z) \big)
  ^{2}\big) z^{N}\exp (-z).  
\end{gather*}
As for $\frac{d}{dz}B(z)$, using the derivation formula \cite{magnus, szego}
\begin{gather*}
  \frac{d}{dz}L_{n}^{N}(z) =-L_{n-1}^{N+1}(z) ,
\end{gather*}
it can be expressed as
\begin{gather}
  \sqrt{2\omega }n\frac{d}{dz}B(z)  = -z^{N}\exp (-z)
    L_{n}^{N}(z) \left( z\frac{d^{2}}{dz^{2}}L_{n}^{N} (
    z ) + ( N+1-z ) \frac{d}{dz}L_{n}^{N}(z) \right)
    \nonumber \\
\hphantom{\sqrt{2\omega }n\frac{d}{dz}B(z)  =}{}-z^{N+1}\exp (-z) \big( L_{n-1}^{N+1}(z) \big)
    ^{2}. \label{b"}
\end{gather}

The dif\/ferential equation satisf\/ied by the Laguerre polynomials is \cite{magnus, szego}
\begin{gather*}
  z\frac{d^{2}}{dz^{2}}L_{n}^{N}(z) + ( N+1-z ) \frac{d}{dz}L_{n}^{N}(z) =-nL_{n}^{N}(z) ,
\end{gather*}
which, inserted in equation~(\ref{b"}), yields
\begin{gather*}
  \sqrt{2\omega }\frac{d}{dz}B(z) =\big( \big( L_{n}^{N} (
  z ) \big) ^{2}-\frac{1}{n}z\big( L_{n-1}^{N+1}(z)
  \big) ^{2}\big) z^{N}\exp (-z).
\end{gather*}
Hence, to satisfy the f\/irst condition (\ref{condb'-IP}), we must choose
\begin{gather*}
  C=\frac{1}{n}.
\end{gather*}

With such a choice for $C$, the second condition~(\ref{condb"-IP}) is then also fulf\/illed because it is obvious that $B(0) = 0$ and, from equation~(\ref{F0}), it results that
\begin{gather*}
  A(0) = F^N_n(0) - \frac{1}{n} F^{N+1}_{n-1}(0) = 0.
\end{gather*}

We conclude that for any $n\ge 1$, the two-step conf\/luent extension $\widetilde{V}_{\rm SUSY}^{(n^{2})}\big(x;\omega ,N,-\Vert \psi_{n}\Vert ^{2}\big)$ satisf\/ies the enlarged shape invariant property~(\ref{ESI-IP}).

Such a property is in particular satisf\/ied if $n=N=1$, as it can be directly verif\/ied. In this case, we indeed have
\begin{gather*}
  L_{0}^{( 2) }(z) =1,\quad L_{1}^{(1) }(z) =2-z,
\end{gather*}
and $C=1$, so that the identity (see equations~(\ref{F}) and~(\ref{Qder}))
\begin{gather*}
  A(z)  =F_{1}^{1}(z) -F_{0}^{2}(z) =
    \frac{1}{\sqrt{2\omega }}\exp (-z) \big(
    Q_{1}^{1}(z)-Q_{0}^{2}(z)\big) \\
\hphantom{A(z)}{}
= \frac{1}{\sqrt{2\omega }}\exp (-z) \left( \sum_{j=0}^{2}\frac{
    d^{j}}{dz^{j}}\big( z^{2}\big) -\sum_{j=0}^{3}\frac{d^{j}}{dz^{j}}\big(
    z ( 2-z ) ^{2}\big) \right)
   = \frac{z^{2} ( 2-z ) }{\sqrt{2\omega }}e^{-z}=B(z)
\end{gather*}
is fulf\/illed, which implies that
\begin{gather*}
  \widetilde{V}_{\rm SUSY}^{( 1^{2}) }\big(x;\omega ,1,- \Vert \psi
  _{1} \Vert ^{2}\big)=\widetilde{V}^{( (0) ^{2})
  }\big(x;\omega ,2,- \Vert \psi _{0} \Vert ^{2}\big)+E_{1} ( \omega
  ) .
\end{gather*}

Let us now turn ourselves to the $n=0$ case. As noted above (see equation~(\ref{no-state})),
the formal eigenfunction
\begin{gather*}
  \widetilde{\psi }_{0}^{(0^{2})}\big(x;\omega ,N,-\Vert \psi
  _{0} \Vert ^{2}\big)=\frac{1}{\Psi _{0}^{(0) }\big(x;\omega
  ,N,- \Vert \psi _{0} \Vert ^{2}\big)},
\end{gather*}
associated to the eigenvalue $E_{0}=0$, is not an eigenstate of $\widetilde{V
}^{(0^{2})}\big(x;\omega ,N,-\Vert \psi _{0}\Vert ^{2}\big)$
and the ground state of this potential is actually given by $\widetilde{\psi }_{1}^{(
0^{2}) }\big(x;\omega ,N,-\Vert \psi _{0}\Vert ^{2}\big)$. The SUSY partner therefore reads
\begin{gather*}
    \widetilde{V}_{\rm SUSY}^{(0^{2})}\big(x;\omega ,N,-\left\Vert \psi
    _{0}\right\Vert ^{2}\big)
 =\widetilde{V}^{(0^{2})}\big(x;\omega
    ,N,- \Vert \psi _{0} \Vert ^{2}\big)-2\big( \log \widetilde{\psi }
    _{1}^{(0^{2})}\big(x;\omega ,N,- \Vert \psi _{0} \Vert
    ^{2}\big)\big) ^{\prime \prime } \\
  \qquad = V(x;\omega,N)
  -2\Big[ \log \big( \psi _{0}(x;\omega ,N)\Psi _{0}^{(0)
    }\big(x;\omega ,N,- \Vert \psi _{0} \Vert ^{2}\big)\widetilde{\psi }
    _{1}^{(0^{2})}\big(x;\omega ,N,- \Vert \psi _{0} \Vert
    ^{2}\big)\big) \Big] ^{\prime \prime },
\end{gather*}
where
\begin{gather*}
  \widetilde{\psi }_{1}^{(0^{2})}\big(x;\omega ,N,- \Vert \psi
  _{0} \Vert ^{2}\big)=-E_{1}(\omega) \psi _{1} ( x;\omega
  ,N ) -\frac{W(\psi _{0},\psi _{1}\,|\, x)}{\Psi _{0}^{(0)
  }\big(x;\omega ,N,- \Vert \psi _{0} \Vert ^{2}\big)}.
\end{gather*}

Using the shape invariance property of~$V$, we can then write
\begin{gather*}
   \widetilde{V}_{\rm SUSY}^{(0^{2})}\big(x;\omega ,N,- \Vert \psi
    _{0} \Vert ^{2}\big) = V ( x;\omega ,N+1 ) +E_{1} ( \omega ) \\
  \qquad {}-2\Big[ \log \big( {-}2\omega \psi _{1}(x;\omega,N) \Psi
    _{0}^{(0) }\big(x;\omega ,N,- \Vert \psi _{0} \Vert
    ^{2}\big)-W(\psi _{0},\psi _{1}\,|\, x)\big) \Big] ^{\prime \prime },
\end{gather*}
that is, with equation~(\ref{seed02}) and (\ref{wron}),
\begin{gather*}
 \widetilde{V}_{\rm SUSY}^{(0^{2})}\big(x;\omega ,N,- \Vert \psi
    _{0} \Vert ^{2}\big)    = V( x;\omega ,N+1) +E_{1}( \omega) - 2 [\log
     \psi _{0}(x;\omega,N)]^{\prime\prime}\\
\qquad\quad {}-2\Big[ \log \big( L_{1}^{N} (
    z ) z^{-(N+1/2)/2}\exp  ( -z/2 ) Q_{0}^{N}(z)-\psi _{0} (
    x;\omega ,N+1 ) \big) \Big] ^{\prime \prime } \\
 \qquad = V^{(0) } ( x;\omega ,N+1 ) +E_{1} ( \omega
    )  \\
\qquad\quad {}     -2\Big\{ \log\big[ z^{-(N+3/2)/2}\exp  ( -z/2 ) \big(
    L_{1}^{N}(z) Q_{0}^{N}(z)-z^{N+1}\big) \big] \Big\}^{\prime \prime }.
\end{gather*}

If there existed some constant $C$ such that
\begin{gather}
  L^N_1(z) Q^N_0(z) - z^{N+1} = C Q^{N+1}_0(z), \label{cond-IP0}
\end{gather}
then we would get
\begin{gather*}
  \widetilde{V}_{\rm SUSY}^{( 0^{2}) }\big(x;\omega ,N,-\Vert \psi
  _{0}\Vert ^{2}\big) = \widetilde{V}^{( 0^{2}) }(x;\omega ,N+1) + E_1(\omega).
\end{gather*}
However, it can be readily seen that equation~(\ref{cond-IP0}) cannot be satisf\/ied by any~$C$, so that we conclude that we do not have any strict nor enlarged shape invariance for the conf\/luent extension~$\widetilde{V}^{( 0^{2}) }$.

\section{Final comments}\label{section6}

Using two-step conf\/luent chains of DBT, we have generated new families of
orthogonal polynomials, associated to novel regular rational extensions of the isotonic and TDPT
potentials, exhibiting an enlarged shape invariance property. Interestingly, in the second case, the ortho\-go\-nal polynomials depend on a free parameter that can be modulated continuously, a feature
already encountered for other extensions based on para-Jacobi polynomials.

Considering chains of arbitrary order and the possibility of obtaining more general families of orthogonal polynomials subjected to multi-parameter dependence would be a very interesting topic for future investigation.

As a f\/inal point, it is worth observing that apart from the Darboux (or dif\/ferential) approach to the construction of (quasi-)isospectral families of potentials, considered in the present paper, there also exist the Abraham--Moses~\cite{abraham} and Pursey~\cite{pursey} (or integral) approaches to the same, coming from the inverse scattering Gel'fand--Levitan or Marchenko technique. Relations between both types of methods have been extensively studied by several authors (see, e.g., \cite{baye, luban, nieto, samsonov95}), showing that they sometimes lead to the same results, but are in general inequivalent. The dif\/ferential procedure being more convenient and easier to use in quantum mechanics has been preferred here. Studying the relation between our results and those of the Abraham--Moses  method would, however, be an interesting open question for future work.

\pdfbookmark[1]{References}{ref}
\LastPageEnding


\begin{thebibliography}{99}
\footnotesize \itemsep=0pt

\bibitem{abraham}
Abraham P.B., Moses H.E., Changes in potentials due to changes in the point
  spectrum: anharmonic oscillators with exact solutions, \href{http://dx.doi.org/10.1103/PhysRevA.22.1333}{\textit{Phys. Rev.~A}}
  \textbf{22} (1980), 1333--1340.

\bibitem{adler-moser}
Adler M., Moser J., On a class of polynomials connected with the {K}orteweg--de
  {V}ries equation, \href{http://dx.doi.org/10.1007/BF01609465}{\textit{Comm. Math. Phys.}} \textbf{61} (1978), 1--30.

\bibitem{adler}
Adler V.{\`E}., On a modif\/ication of {C}rum's method, \href{http://dx.doi.org/10.1007/BF01035458}{\textit{Theoret. and
  Math. Phys.}} \textbf{101} (1994), 1381--1386.

\bibitem{BGQ}
Bagchi B., Grandati Y., Quesne C., Rational extensions of the trigonometric
  {D}arboux--{P}\"oschl--{T}eller potential based on para-{J}acobi polynomials,
  \href{http://dx.doi.org/10.1063/1.4922017}{\textit{J.~Math. Phys.}} \textbf{56} (2015), 062103, 11~pages,
  \href{http://arxiv.org/abs/1411.7857}{arXiv:1411.7857}.

\bibitem{baye}
Baye D., Phase-equivalent potentials for arbitrary modif\/ications of the bound
  spectrum, \href{http://dx.doi.org/10.1103/PhysRevA.48.2040}{\textit{Phys. Rev.~A}} \textbf{48} (1993), 2040--2047.

\bibitem{berezin}
Berezin F.A., Shubin M.A., The Schr\"{o}dinger equation, \href{http://dx.doi.org/10.1007/978-94-011-3154-4}{Kluwer}, Dordrecht,
  1991.

\bibitem{fernandez2}
Bermudez D., Fern{\'a}ndez~C. D.J., Fern{\'a}ndez-Garc{\'{\i}}a N., Wronskian
  dif\/ferential formula for conf\/luent supersymmetric quantum mechanics,
  \href{http://dx.doi.org/10.1016/j.physleta.2011.12.020}{\textit{Phys. Lett.~A}} \textbf{376} (2012), 692--696, \href{http://arxiv.org/abs/1109.0079}{arXiv:1109.0079}.

\bibitem{burchnall}
Burchnall J.L., Chaundy T.W., A set of dif\/ferential equations which can be
  solved by polynomials, \href{http://dx.doi.org/10.1112/plms/s2-30.1.401}{\textit{Proc. London Math. Soc.}} \textbf{S2-30}
  (1930), 401--414.

\bibitem{calogero}
Calogero F., Yi G., Can the {\it general} solution of the second-order {ODE}
  characterizing {J}acobi polynomials be {\it polynomial}?, \href{http://dx.doi.org/10.1088/1751-8113/45/9/095206}{\textit{J.~Phys.~A:
  Math. Theor.}} \textbf{45} (2012), 095206, 4~pages.

\bibitem{Ramos}
Cari{\~n}ena J.F., Ramos A., Integrability of the {R}iccati equation from a
  group-theoretical viewpoint, \href{http://dx.doi.org/10.1142/S0217751X9900097X}{\textit{Internat.~J. Modern Phys.~A}} \textbf{14}
  (1999), 1935--1951, \href{http://arxiv.org/abs/math-ph/9810005}{math-ph/9810005}.

\bibitem{carinena2}
Cari{\~n}ena J.F., Ramos A., Fern{\'a}ndez~C. D.J., Group theoretical approach
  to the intertwined {H}amiltonians, \href{http://dx.doi.org/10.1006/aphy.2001.6179}{\textit{Ann. Physics}} \textbf{292} (2001),
  42--66, \href{http://arxiv.org/abs/math-ph/0311029}{math-ph/0311029}.

\bibitem{contreras}
Contreras-Astorga A., Fern{\'a}ndez~C. D.J., Supersymmetric partners of the
  trigonometric {P}\"oschl--{T}eller potentials, \href{http://dx.doi.org/10.1088/1751-8113/41/47/475303}{\textit{J.~Phys.~A: Math.
  Theor.}} \textbf{41} (2008), 475303, 18~pages, \href{http://arxiv.org/abs/0809.2760}{arXiv:0809.2760}.

\bibitem{cooper}
Cooper F., Khare A., Sukhatme U., Supersymmetry in quantum mechanics, \href{http://dx.doi.org/10.1142/9789812386502}{World
  Scientif\/ic Publishing Co.}, Inc., River Edge, NJ, 2001.

\bibitem{crum}
Crum M.M., Associated {S}turm--{L}iouville systems, \href{http://dx.doi.org/10.1093/qmath/6.1.121}{\textit{Quart.~J. Math.
  Oxford Ser.~(2)}} \textbf{6} (1955), 121--127, \href{http://arxiv.org/abs/physics/9908019}{physics/9908019}.

\bibitem{darboux2}
Darboux G., Sur une proposition relative aux \'equations lin\'eaires,
  \textit{C.~R.~Acad. Sci. Paris} \textbf{94} (1882), 1456--1459.

\bibitem{darboux}
Darboux G., Le\c{c}ons sur la th{\'e}orie g{\'e}n{\'e}rale des surfaces et les
  applications g{\'e}om{\'e}triques du calcul inf\/init{\'e}simal.~II, 2nd ed.,
  Gauthier-Villars, Paris, 1915.

\bibitem{Duran}
Dur{\'a}n A.J., P{\'e}rez M., Admissibility condition for exceptional
  {L}aguerre polynomials, \href{http://dx.doi.org/10.1016/j.jmaa.2014.11.035}{\textit{J.~Math. Anal. Appl.}} \textbf{424} (2015),
  1042--1053, \href{http://arxiv.org/abs/1409.4901}{arXiv:1409.4901}.

\bibitem{magnus}
Erd\'{e}lyi A., Magnus W., Oberhettinger F., Tricomi F.G., Higher
  transcendental functions, McGraw-Hill, New York, 1953.

\bibitem{fernandez03}
Fern{\'a}ndez~C. D.J., Salinas-Hern{\'a}ndez E., The conf\/luent algorithm in
  second-order supersymmetric quantum mechanics, \href{http://dx.doi.org/10.1088/0305-4470/36/10/313}{\textit{J.~Phys.~A: Math.
  Gen.}} \textbf{36} (2003), 2537--2543, \href{http://arxiv.org/abs/quant-ph/0303123}{quant-ph/0303123}.

\bibitem{fernandez}
Fern{\'a}ndez~C. D.J., Salinas-Hern{\'a}ndez E., Wronskian formula for
  conf\/luent second-order supersymmetric quantum mechanics, \href{http://dx.doi.org/10.1016/j.physleta.2005.02.020}{\textit{Phys.
  Lett.~A}} \textbf{338} (2005), 13--18, \href{http://arxiv.org/abs/quant-ph/0502147}{quant-ph/0502147}.

\bibitem{fernandez3}
Fern{\'a}ndez~C. D.J., Salinas-Hern{\'a}ndez E., Hyperconf\/luent third-order
  supersymmetric quantum mechanics, \href{http://dx.doi.org/10.1088/1751-8113/44/36/365302}{\textit{J.~Phys.~A: Math. Theor.}}
  \textbf{44} (2011), 365302, 11~pages, \href{http://arxiv.org/abs/1105.2333}{arXiv:1105.2333}.

\bibitem{GGM}
G\'omez-Ullate D., Grandati Y., Milson R., Extended Krein--Adler theorem for
  the translationally shape invariant potentials, \href{http://dx.doi.org/10.1063/1.4871443}{\textit{J.~Math. Phys.}}
  \textbf{55} (2014), 043510, 30~pages, \href{http://arxiv.org/abs/1309.3756}{arXiv:1309.3756}.

\bibitem{gomez3}
G{\'o}mez-Ullate D., Kamran N., Milson R., An extended class of orthogonal
  polynomials def\/ined by a {S}turm--{L}iouville problem, \href{http://dx.doi.org/10.1016/j.jmaa.2009.05.052}{\textit{J.~Math. Anal.
  Appl.}} \textbf{359} (2009), 352--367, \href{http://arxiv.org/abs/0807.3939}{arXiv:0807.3939}.

\bibitem{grandati4}
Grandati Y., Solvable rational extensions of the isotonic oscillator,
  \href{http://dx.doi.org/10.1016/j.aop.2011.03.001}{\textit{Ann. Physics}} \textbf{326} (2011), 2074--2090, \href{http://arxiv.org/abs/1101.0055}{arXiv:1101.0055}.

\bibitem{grandati5}
Grandati Y., Multistep {DBT} and regular rational extensions of the isotonic
  oscillator, \href{http://dx.doi.org/10.1016/j.aop.2012.07.004}{\textit{Ann. Physics}} \textbf{327} (2012), 2411--2431,
  \href{http://arxiv.org/abs/1108.4503}{arXiv:1108.4503}.

\bibitem{grandati6}
Grandati Y., New rational extensions of solvable potentials with f\/inite bound
  state spectrum, \href{http://dx.doi.org/10.1016/j.physleta.2012.09.037}{\textit{Phys. Lett.~A}} \textbf{376} (2012), 2866--2872,
  \href{http://arxiv.org/abs/1203.4149}{arXiv:1203.4149}.

\bibitem{grandati2}
Grandati Y., A short proof of the {G}aillard--{M}atveev theorem based on shape
  invariance arguments, \href{http://dx.doi.org/10.1016/j.physleta.2014.03.020}{\textit{Phys. Lett.~A}} \textbf{378} (2014), 1755--1759,
  \href{http://arxiv.org/abs/1211.2392}{arXiv:1211.2392}.

\bibitem{grandati}
Grandati Y., B{\'e}rard A., Rational solutions for the
  {R}iccati--{S}chr\"odinger equations associated to translationally shape
  invariant potentials, \href{http://dx.doi.org/10.1016/j.aop.2010.03.008}{\textit{Ann. Physics}} \textbf{325} (2010), 1235--1259,
  \href{http://arxiv.org/abs/0910.4810}{arXiv:0910.4810}.

\bibitem{grandatiDPT}
Grandati Y., B{\'e}rard A., Comments on the generalized {SUSY} {QM} partnership
  for {D}arboux--{P}\"oschl--{T}eller potential and exceptional {J}acobi
  polynomials, \href{http://dx.doi.org/10.1007/s10665-012-9601-x}{\textit{J.~Engrg. Math.}} \textbf{82} (2013), 161--171.

\bibitem{hartman}
Hartman P., Ordinary dif\/ferential equations, John Wiley \& Sons, Inc., New
  York~-- London~-- Sydney, 1964.

\bibitem{keung}
Keung W.-Y., Sukhatme U.P., Wang Q.M., Imbo T.D., Families of strictly
  isospectral potentials, \href{http://dx.doi.org/10.1088/0305-4470/22/21/002}{\textit{J.~Phys.~A: Math. Gen.}} \textbf{22} (1989),
  L987--L992.

\bibitem{krein}
Krein M.G., On a continual analogue of a {C}hristof\/fel formula from the theory
  of orthogonal polynomials, \textit{Dokl. Akad. Nauk SSSR} \textbf{113}
  (1957), 970--973.

\bibitem{luban}
Luban M., Pursey D.L., New {S}chr\"odinger equations for old: inequivalence of
  the {D}arboux and {A}braham--{M}oses constructions, \href{http://dx.doi.org/10.1103/PhysRevD.33.431}{\textit{Phys. Rev.~D}}
  \textbf{33} (1986), 431--436.

\bibitem{matveev}
Matveev V.B., Generalized {W}ronskian formula for solutions of the {K}d{V}
  equations: f\/irst applications, \href{http://dx.doi.org/10.1016/0375-9601(92)90362-P}{\textit{Phys. Lett.~A}} \textbf{166} (1992),
  205--208.

\bibitem{matveev2}
Matveev V.B., Positons: slowly decreasing analogues of solitons,
  \href{http://dx.doi.org/10.1023/A:1015149618529}{\textit{Theoret. and Math. Phys.}} \textbf{131} (2002), 483--497.

\bibitem{messiah}
Messiah A., M\'ecanique quantique, Vol.~1, Dunod, Paris, 1959.

\bibitem{mielnik}
Mielnik B., Nieto L.M., Rosas-Ortiz O., The f\/inite dif\/ference algorithm for
  higher order supersymmetry, \href{http://dx.doi.org/10.1016/S0375-9601(00)00226-7}{\textit{Phys. Lett.~A}} \textbf{269} (2000),
  70--78, \href{http://arxiv.org/abs/quant-ph/0004024}{quant-ph/0004024}.

\bibitem{nieto}
Nieto M.M., Relationship between supersymmetry and the inverse method in
  quantum mechanics, \href{http://dx.doi.org/10.1016/0370-2693(84)90339-3}{\textit{Phys. Lett.~B}} \textbf{145} (1984), 208--210.

\bibitem{odake}
Odake S., Sasaki R., Inf\/initely many shape invariant potentials and new
  orthogonal polynomials, \href{http://dx.doi.org/10.1016/j.physletb.2009.08.004}{\textit{Phys. Lett.~B}} \textbf{679} (2009), 414--417,
  \href{http://arxiv.org/abs/0906.0142}{arXiv:0906.0142}.

\bibitem{pursey}
Pursey D.L., New families of isospectral {H}amiltonians, \href{http://dx.doi.org/10.1103/PhysRevD.33.1048}{\textit{Phys. Rev.~D}}
  \textbf{33} (1986), 1048--1055.

\bibitem{quesne1}
Quesne C., Exceptional orthogonal polynomials, exactly solvable potentials and
  supersymmetry, \href{http://dx.doi.org/10.1088/1751-8113/41/39/392001}{\textit{J.~Phys.~A: Math. Theor.}} \textbf{41} (2008), 392001,
  6~pages, \href{http://arxiv.org/abs/0807.4087}{arXiv:0807.4087}.

\bibitem{quesne}
Quesne C., Solvable rational potentials and exceptional orthogonal polynomials
  in supersymmetric quantum mechanics, \href{http://dx.doi.org/10.3842/SIGMA.2009.084}{\textit{SIGMA}} \textbf{5} (2009), 084,
  24~pages, \href{http://arxiv.org/abs/0906.2331}{arXiv:0906.2331}.

\bibitem{quesne4}
Quesne C., Higher-order {SUSY}, exactly solvable potentials, and exceptional
  orthogonal polynomials, \href{http://dx.doi.org/10.1142/S0217732311036383}{\textit{Modern Phys. Lett.~A}} \textbf{26} (2011),
  1843--1852, \href{http://arxiv.org/abs/1106.1990}{arXiv:1106.1990}.

\bibitem{quesne6}
Quesne C., Novel enlarged shape invariance property and exactly solvable
  rational extensions of the {R}osen--{M}orse~{II} and {E}ckart potentials,
  \href{http://dx.doi.org/10.3842/SIGMA.2012.080}{\textit{SIGMA}} \textbf{8} (2012), 080, 19~pages, \href{http://arxiv.org/abs/1208.6165}{arXiv:1208.6165}.

\bibitem{quesne7}
Quesne C., Revisiting (quasi-)exactly solvable rational extensions of the
  {M}orse potential, \href{http://dx.doi.org/10.1142/S0217751X1250073X}{\textit{Internat.~J. Modern Phys.~A}} \textbf{27} (2012),
  1250073, 18~pages, \href{http://arxiv.org/abs/1203.1811}{arXiv:1203.1811}.

\bibitem{samsonov95}
Samsonov B.F., On the equivalence of the integral and the dif\/ferential exact
  solution generation methods for the one-dimensional {S}chr\"odinger equation,
  \href{http://dx.doi.org/10.1088/0305-4470/28/23/036}{\textit{J.~Phys.~A: Math. Gen.}} \textbf{28} (1995), 6989--6998.

\bibitem{samsonov}
Samsonov B.F., New possibilities for supersymmetry breakdown in quantum
  mechanics and second-order irreducible {D}arboux transformations,
  \href{http://dx.doi.org/10.1016/S0375-9601(99)00736-7}{\textit{Phys. Lett.~A}} \textbf{263} (1999), 274--280,
  \href{http://arxiv.org/abs/quant-ph/9904009}{quant-ph/9904009}.

\bibitem{schulze}
Schulze-Halberg A., Wronskian representation for conf\/luent supersymmetric
  transformation chains of arbitrary order, \href{http://dx.doi.org/10.1140/epjp/i2013-13068-2}{\textit{Eur. Phys.~J. Plus}}
  \textbf{128} (2013), 69, 17~pages.

\bibitem{sparenberg}
Sparenberg J.-M., Baye D., Supersymmetric transformations of real potentials on
  the line, \href{http://dx.doi.org/10.1088/0305-4470/28/17/033}{\textit{J.~Phys.~A: Math. Gen.}} \textbf{28} (1995), 5079--5095.

\bibitem{szego}
Szeg{\H{o}} G., Orthogonal polynomials, \textit{American Mathematical Society
  Colloquium Publications}, Vol.~23, 4th~ed., Amer. Math. Soc., Providence,
  R.I., 1975.

\bibitem{vein}
Vein R., Dale P., Determinants and their applications in mathematical physics,
  \href{http://dx.doi.org/10.1007/b98968}{\textit{Applied Mathematical Scien\-ces}}, Vol.~134, Springer-Verlag, New York,
  1999.

\end{thebibliography}
\end{document}